\documentclass[fleqn,usenatbib]{mnras}

\usepackage{newtxtext,newtxmath}
\usepackage[T1]{fontenc}

\DeclareRobustCommand{\VAN}[3]{#2}
\let\VANthebibliography\thebibliography
\def\thebibliography{\DeclareRobustCommand{\VAN}[3]{##3}\VANthebibliography}

\usepackage{graphicx}	% Including figure files
\usepackage{amsmath}	% Advanced maths commands
\usepackage{xspace}
\usepackage{comment}
\usepackage{gensymb}
\usepackage[export]{adjustbox}
\usepackage{physics}
\usepackage{bm}

\usepackage{siunitx}

\usepackage{threeparttable} 
\newlength{\abovecaptionskip}%
\newcommand{\Msun}{\text{M}_\odot}

\newcommand{\Mpc}{\text{Mpc}}
\newcommand{\Gpc}{\text{Gpc}}

\newcommand{\eV}{\text{eV}}

\newcommand{\bahamas}{BAHAMAS}
\newcommand{\flamingo}{FLAMINGO}
\newcommand{\swift}{\textsc{Swift}\xspace}

\newcommand{\healpix}{\textsc{HEALPix}\xspace}
\newcommand{\namaster}{\textsc{NaMaster}\xspace}

\newcommand{\panphasia}{\textsc{panphasia}\xspace}
\newcommand{\monofonIC}{\textsc{monofonIC}\xspace}

\newcommand{\diff}{\mathrm{d}}

\title[FLAMINGO: Baryons and the $S_8$ tension]{The FLAMINGO project: revisiting the $S_8$ tension and the role of baryonic physics}

\author[I. G. McCarthy et al.]{Ian G. McCarthy,$^{1}$\thanks{E-mail: i.g.mccarthy@ljmu.ac.uk}
Jaime Salcido,$^{1}$
Joop Schaye,$^{2}$
Juliana Kwan,$^{1}$
Willem Elbers,$^{3}$
Roi Kugel,$^{2}$ \newauthor
Matthieu Schaller,$^{4,2}$
John C. Helly,$^{3}$
Joey Braspenning,$^{2}$
Carlos S. Frenk,$^{3}$
Marcel P. van Daalen,$^{2}$ \newauthor
Bert Vandenbroucke,$^{2}$
Jonah T. Conley,$^{1}$
Andreea S. Font,$^{1}$
Amol Upadhye,$^{1}$
\\
$^{1}$Astrophysics Research Institute, Liverpool John Moores University, Liverpool, L3 5RF, UK\\
$^{2}$Leiden Observatory, Leiden University, PO Box 9513, 2300 RA Leiden, the Netherlands\\
$^{3}$Institute for Computational Cosmology, Department of Physics, University of Durham, South Road, Durham, DH1 3LE, UK\\
$^{4}$Lorentz Institute for Theoretical Physics, Leiden University, PO box 9506, 2300 RA Leiden, the Netherlands
}

\date{Accepted XXX. Received YYY; in original form ZZZ}

\pubyear{2015}

\begin{document}
\label{firstpage}
\pagerange{\pageref{firstpage}--\pageref{lastpage}}
\maketitle

% Abstract of the paper
\begin{abstract}
A number of recent studies have found evidence for a tension between observations of large-scale structure (LSS) and the predictions of the standard model of cosmology with the cosmological parameters fit to the cosmic microwave background (CMB).  The origin of this `$S_8$ tension' remains unclear, but possibilities include new physics beyond the standard model, unaccounted for systematic errors in the observational measurements and/or uncertainties in the role that baryons play.  Here we carefully examine the latter possibility using the new \flamingo\ suite of large-volume cosmological hydrodynamical simulations.  We project the simulations onto observable harmonic space and compare with observational measurements of the power and cross-power spectra of cosmic shear, CMB lensing, and the thermal Sunyaev-Zel'dovich (tSZ) effect.  We explore the dependence of the predictions on box size and resolution, cosmological parameters including the neutrino mass, and the efficiency and nature of baryonic `feedback'.  Despite the wide range of astrophysical behaviours simulated, we find that baryonic effects are not sufficiently large to remove the $S_8$ tension.   Consistent with recent studies, we find the CMB lensing power spectrum is in excellent agreement with the standard model, whilst the cosmic shear power spectrum, tSZ effect power spectrum, and the cross-spectra between shear, CMB lensing, and the tSZ effect are all in varying degrees of tension with the CMB-specified standard model.  These results suggest that some mechanism is required to slow the growth of fluctuations at late times and/or on non-linear scales, but that it is unlikely that baryon physics is driving this modification.   
\end{abstract}

% Select between one and six entries from the list of approved keywords.
% Don't make up new ones.

%%%%%%%%%%%%%%%%%%%%%%%%%%%%%%%%%%%%%%%%%%%%%%%%%%

%%%%%%%%%%%%%%%%% BODY OF PAPER %%%%%%%%%%%%%%%%%%

\section{Introduction}
\label{sec:intro}

The standard model of cosmology, the so-called $\Lambda$CDM model, is based on the Friedmann–Lemaître–Robertson–Walker solution to Einstein's field equations for an isotropic and homogeneous universe.  The standard model contains only 6 free parameters but describes a wealth of large-scale cosmological data remarkably well, including the temperature and polarization anisotropies in the cosmic microwave background (CMB), measurements of baryon acoustic oscillations (BAO) in the clustering of galaxies, the redshift--distance relations of supernovae (Sn) Type Ia, and measurements of the growth of large-scale structure (LSS) including the abundance of galaxy clusters, galaxy clustering, cosmic shear and CMB lensing, and the thermal Sunyaev-Zel'dovich (tSZ) effect (see, e.g., \citealt{Planck2020cosmopars}).  Fits to these data point to a Universe which is spatially flat and whose present-day energy density is dominated by dark matter and dark energy.  However, the physical nature of these components has so far remained elusive.  

A fruitful avenue for exploration into the nature of dark matter and dark energy, and to test the standard cosmological paradigm generally, is to look for signs of deviations in cosmological data sets from the predictions of the standard model and its possible extensions.  On this front there has been much activity in the past few years as the fidelity of cosmological data sets has rapidly increased.  Interestingly, even though the standard model describes many cosmological data sets extremely well, the best-fit parameter values from different observables do not always appear fully consistent with each other.  The most notable example of this is the so-called `Hubble tension', which is that measurements of the local expansion rate of space yield a value for Hubble's constant, $H_0$, that is larger than predicted by the standard model of cosmology when it is fit to the CMB temperature and polarization anisotropies and BAO data (e.g., \citealt{Planck2020cosmopars,Riess2022}).  This tension has now reached the $\approx 5\sigma$ level \citep{Riess2022}.  

Another notable tension, which we focus on in the present study, is the so-called `$S_8$ tension', where $S_8$ is defined as $\sigma_8 \sqrt{\Omega_\textrm{m}/0.3}$, where $\Omega_m$ is the present-day matter density and $\sigma_8$ is the linearly-evolved variance of the present-day matter density field filtered on a $8$ Mpc/$h$ scale.  Note that $S_8$ best describes the combination of $\sigma_8$ and $\Omega_m$ that is constrained by cosmic shear (weak lensing) data.  In this case, the best-fit value of $S_8$ from several low-redshift observations of LSS, including cosmic shear (e.g., \citealt{Heymans2021,Abbott2022}) and other probes, appears to be in mild ($\approx1$-$3\sigma$) tension with the predictions of the standard model with parameter values specified by the CMB and BAO and, as we discuss below, also CMB lensing.  While a tension of this magnitude is often not regarded as being statistically compelling, it is worth highlighting here that the tension has been persistent for nearly a decade now, since the first Planck data release, and spans several independent probes, each of which appear to show tensions with this level of significance and in the same direction (e.g., see figure 1 of \citealt{McCarthy2018}).

Various possible solutions have been put forward to reconcile the low-redshift LSS observations with the primary CMB + BAO combination.  This includes mischaracterised systematic uncertainties in the LSS observations (e.g., photometric redshift, galaxy shape, and intrinsic alignment uncertainties in cosmic shear measurements, non-linear biasing in galaxy clustering, halo mass biasing in cluster counts, etc.), or possibly even in the primary CMB measurements (e.g., \citealt{Addison2016,Planck2017}; but see \citealt{Rosenberg2022}).  On the theory side, LSS tests of cosmology often probe the non-linear regime and therefore require cosmological simulations, or models which have been calibrated on such simulations, to predict the clustering of matter on small scales and at late times.  Furthermore, part and parcel of this non-linear evolution is that matter collapses along filaments forming `haloes' at the nodes.  Here the densities reach sufficiently high values that radiative cooling of the gas becomes efficient, leading to further collapse and eventually galaxy formation (e.g., \citealt{White1991}).  With this comes a variety of energetic feedback processes associated with the formation of stars and the accretion of matter onto supermassive black holes.  

It is straightforward to show that the energy released by the accretion of matter onto supermassive black holes can be of cosmological significance if it is able to efficiently couple to the gas (e.g., \citealt{Silk1998}).  That is, the energy is sufficient to expel baryons from galaxy groups (e.g., \citealt{McCarthy2010,McCarthy2011}), which will also lead to a back reaction on the dark matter halo (e.g., \citealt{VanDaalen2011}).  In short, in the presence of energetic feedback, we expect the clustering of matter to be significantly affected on non-linear scales, and the results of full cosmological hydrodynamical simulations (e.g., \citealt{VanDaalen2011,VanDaalen2020,Mummery2017,Springel2018,Salcido2023}) as well as analytic halo models that use the observed baryon content of massive haloes as input (e.g., \citealt{Debackere2020}) back up this physical intuition.  Recent studies have shown that if such processes are not accounted for, they will lead to significant biases in the recovered cosmological parameters in forthcoming surveys (e.g., \citealt{Semboloni2011,Semboloni2013,Schneider2020,Castro2021}; for a recent review see \citealt{Chisari2019}).

But what role, if any, do unaccounted for (or mischaracterised) baryonic effects have on the current $S_8$ tension?  Based on the BAHAMAS simulations \citep{McCarthy2017}, we have previously argued that the effects are likely to be too small to explain the current tension \citep{McCarthy2018}.   Consistent with this are the findings of several recent analyses of cosmic shear data, which have made marginal detections of the impact of baryons on the matter clustering and find that its magnitude aligns well with the predictions of cosmological hydrodynamical simulations such as \bahamas\ (see, e.g, \citealt{Troster2022,Chen2023,Arico2023}).  However, \citet{Amon2022} and \citet{Preston2023} have recently challenged this conventional wisdom, showing that the existing constraints on the impact of baryons are sensitive to the adopted priors on the baryon parameters and that with a wider set of priors, allowing for much more aggressive feedback beyond what is typically simulated, it may be possible after all to reconcile the primary CMB(+BAO+CMB lensing) measurements with low-redshift LSS measurements.

In the present study we revisit the $S_8$ tension and the role that baryons play.  We use the new \flamingo\ suite of large-volume cosmological simulations \citep{Schaye2023,Kugel2023} that includes variations in box size, resolution, cosmology (including massive neutrino cosmologies), and, importantly, a careful, systematic variation of the efficiencies and nature of feedback from star formation and Active Galactic Nuclei (AGN).  We project the simulations onto observable harmonic space and make predictions for the power and cross-power spectra of cosmic shear, CMB lensing, and the thermal Sunyaev-Zel'dovich (tSZ) effect, which sample fluctuations over a very wide range of physical scales and redshifts and have different sensitivities to halo mass.  While there is a significant degree of overlap in terms of the scales between the various auto- and cross-power spectra (allowing for important consistency checks), there is also a great deal of complementarity.  We compare the \flamingo\ predictions with the latest measurements of these quantities from the KiDS 1000 \citep{Troster2022} and DES Y3 survey \citep{Doux2022} cosmic shear data, Planck and SPT tSZ data (\citealt{Bolliet2018} and \citealt{Reichardt2021}, respectively), and Planck, ACT, and SPT CMB lensing data (\citealt{Planck_lensing}, \citealt{Wu2019}, and \citealt{Qu2023}, respectively).  Despite the wide range of astrophysical behaviours simulated, we will show that baryonic effects and their uncertainties encapsulated within the \flamingo\ simulations are not sufficiently large to significantly alter the current $S_8$ discussion.  

The present study is organised as follows.  In Section \ref{sec:sims_and_obs} we describe the \flamingo\ suite of cosmological hydrodynamical simulations and our approach to projecting these simulations onto observable harmonic space.  In Section \ref{sec:results} we present our main results, including an examination of the box size, resolution, cosmological, and feedback dependencies of the auto- and cross-spectra involving cosmic shear, CMB lensing, and tSZ effect data.  We also compare to the most recent measurements of these quantities.  In Section \ref{sec:conclusions} we summarise our main findings and conclude.

\section{Simulations and computation of cosmological observables}
\label{sec:sims_and_obs}

\subsection{Description of simulations}
\label{sec:sims}

We provide here a brief summary of the \flamingo\ simulations, referring the reader to \citet{Schaye2023} and \citet{Kugel2023} for in depth presentations.

\begin{table*}
	\centering
	\caption{\flamingo\ hydrodynamical simulations. The first four lines list the simulations that use the fiducial galaxy formation model and assume the fiducial cosmology (D3A), but use different volumes and resolutions. The remaining lines list the model variations, which all use a 1~Gpc box and intermediate resolution. The columns list the simulation identifier (where m8, m9 and m10 indicate $\log_{10}$ of the mean initial baryonic particle mass and correspond to high, intermediate, and low resolution, respectively; absence of this part implies m9 resolution); the number of standard deviations by which the observed stellar masses are shifted before calibration, $\Delta m_\ast$; the number of standard deviations by which the observed cluster gas fractions are shifted before calibration, $\Delta f_\text{gas}$; the AGN feedback implementation (thermal or jets); the comoving box side length, $L$; the number of baryonic particles, $N_\text{b}$ (which equals the number of CDM particles, $N_\text{CDM})$; the number of neutrino particles, $N_\nu$; the initial mean baryonic particle mass, $m_\text{g}$; the mean CDM particle mass, $m_\text{CDM}$; the Plummer-equivalent comoving gravitational softening length, $\epsilon_\text{com}$; the maximum proper gravitational softening length, $\epsilon_\text{prop}$; and the assumed cosmology which is specified in Table~\ref{tab:cosmologies}.}
	\label{tab:simulations}
	\begin{tabular}{lrrllrrllrrl} 
		\hline
		Identifier & $\Delta m_\ast$ & $\Delta f_\text{gas}$ & AGN & $L$ & $N_\text{b}$ & $N_\nu$ & $m_\text{g}$ & $m_\text{CDM}$ & $\epsilon_\text{com}$ & $\epsilon_\text{prop}$ & Cosmology \\ 
		           & ($\sigma$) & ($\sigma$) && (cGpc) &&& ($\Msun$) & ($\Msun$)  & (ckpc) & (pkpc) \\
		\hline
		L1\_m8             & 0 & 0 & thermal & 1 & $3600^3$ & $2000^3$ & $1.34\times 10^8$ & $7.06\times 10^8$    & 11.2  & 2.85 & D3A \\
		L1\_m9             & 0 & 0 & thermal & 1 & $1800^3$ & $1000^3$ & $1.07\times 10^9$ & $5.65\times 10^9$    & 22.3  & 5.70 & D3A \\
		L1\_m10              & 0 & 0 & thermal & 1 & $900^3$ & $500^3$ & $8.56\times 10^9$ & $4.52\times 10^{10}$ & 44.6 & 11.40 & D3A \\
		L2p8\_m9           & 0 & 0 & thermal &2.8& $5040^3$ & $2800^3$ & $1.07\times 10^9$ & $5.65\times 10^9$    & 22.3  & 5.70 & D3A \\
		fgas$+2\sigma$  & 0 & $+2$ & thermal & 1 & $1800^3$ & $1000^3$ & $1.07\times 10^9$ & $5.65\times 10^9$ & 22.3  & 5.70 & D3A \\
		fgas$-2\sigma$  & 0 & $-2$ & thermal & 1 & $1800^3$ & $1000^3$ & $1.07\times 10^9$ & $5.65\times 10^9$    & 22.3  & 5.70 & D3A \\
		fgas$-4\sigma$  & 0 & $-4$ & thermal & 1 & $1800^3$ & $1000^3$ & $1.07\times 10^9$ & $5.65\times 10^9$    & 22.3  & 5.70 & D3A \\
		fgas$-8\sigma$  & 0 & $-8$ & thermal & 1 & $1800^3$ & $1000^3$ & $1.07\times 10^9$ & $5.65\times 10^9$    & 22.3  & 5.70 & D3A \\
		M*$-\sigma$ & $-1$ & 0 & thermal & 1 & $1800^3$ & $1000^3$ & $1.07\times 10^9$ & $5.65\times 10^9$    & 22.3  & 5.70 & D3A \\
		M*$-\sigma$\_fgas$-4\sigma$     & $-1$ & $-4$ & thermal & 1 & $1800^3$ & $1000^3$ & $1.07\times 10^9$ & $5.65\times 10^9$    & 22.3  & 5.70 & D3A \\
		Jet             & 0 & 0 & jets & 1 & $1800^3$ & $1000^3$ & $1.07\times 10^9$ & $5.65\times 10^9$    & 22.3  & 5.70 & D3A \\
		Jet\_fgas$-4\sigma$  & 0 & $-4$ & jets & 1 & $1800^3$ & $1000^3$ & $1.07\times 10^9$ & $5.65\times 10^9$    & 22.3  & 5.70 & D3A \\
		Planck          & 0 & 0 & thermal & 1 & $1800^3$ & $1000^3$  & $1.07\times 10^9$ & $5.72\times 10^9$    & 22.3  & 5.70 & Planck\\
		PlanckNu0p24Var & 0 & 0 & thermal & 1 & $1800^3$ & $1000^3$ & $1.06\times 10^9$ & $5.67\times 10^9$    & 22.3  & 5.70 & PlanckNu0p24Var \\
		PlanckNu0p24Fix & 0 & 0 & thermal & 1 & $1800^3$ & $1000^3$ & $1.07\times 10^9$ & $5.62\times 10^9$    & 22.3  & 5.70 & PlanckNu0p24Fix \\
		LS8             & 0 & 0 & thermal & 1 & $1800^3$ & $1000^3$ & $1.07\times 10^9$ & $5.65\times 10^9$    & 22.3  & 5.70 & LS8 \\
		\hline
	\end{tabular}
\end{table*}

The \flamingo\ simulations were performed using \swift \citep{SWIFT_release}, a fully open-source coupled cosmology, gravity, hydrodynamics, and galaxy formation code\footnote{Publicly available, including the version used for these simulations, at \url{www.swiftsim.com}}.  Gravitational forces are computed using a 4$^{\mathrm{th}}$-order fast multipole method \citep{Greengard1987, Cheng1999, Dehnen2014} on small scales and a particle-mesh method solved in Fourier space on large scales, following the force splitting approach of \cite{Bagla2003}.  The hydrodynamic equations are solved using the smoothed-particle hydrodynamics (SPH) method \citep[for a review, see][]{Price2012}, in particular the SPHENIX flavour of SPH \citep{Borrow2022sphenix} which was designed specifically for simulations of galaxy formation.

The suite consists of the 16 hydrodynamical simulations, listed in Table~\ref{tab:simulations} (reproduced from Table 2 of \citealt{Schaye2023}), and 12 gravity-only simulations.  We examine only the hydro simulations in the present study.  The majority of the runs adopt a $(1~\Gpc)^3$ cubic volume, denoted by `L1' in the simulation identifier, although one run has a volume of $(2.8~\Gpc)^3$ (`L2p8'). The simulations span three different resolutions (`m10', `m9' and `m8', where the number indicates the rounded logarithm base 10 of the baryonic particle mass), with the mass (spatial) resolution between consecutive resolutions changing by a factor of 8 (2).  Most runs adopt intermediate resolution (`m9'), which corresponds to an (initial) mean baryonic particle mass of $\approx 1\times 10^9\,\Msun$, a mean cold dark matter particle mass of $\approx 6\times 10^9\,\Msun$, and a maximum proper gravitational softening length of 5.7~kpc, all of which are similar to our previous BAHAMAS simulations \citep{McCarthy2017} but within considerably larger volumes than BAHAMAS.  At $z>2.91$ the softening length is held constant in comoving units at 22.3~kpc. All runs use equal numbers of baryonic and dark matter particles, while the number of neutrino particles is a factor $1.8^3$ smaller. Table~\ref{tab:simulations} provides the parameter values specifying the numerical resolution the various runs.

The values of the cosmological parameters for our fiducial model are the maximum likelihood values from the Dark Energy Survey Year Three (DES Y3; \citealt{Abbott2022}) `3×2pt + All Ext.' $\Lambda$CDM cosmology (`D3A' in Table~\ref{tab:cosmologies}). These values assume a spatially flat universe and are based on the combination of constraints from DES~Y3 `$3\times2$-point' correlation functions: cosmic shear, galaxy clustering, and galaxy-galaxy lensing, with constraints from external data from BAO, redshift-space distortions, SN Type Ia, and Planck observations of the CMB (including CMB lensing), Big Bang nucleosynthesis, and local measurements of the Hubble constant (see \citealt{Abbott2022} for details). Our fiducial cosmology, D3A, uses the minimum neutrino mass allowed by neutrino oscillation experiments of $\sum m_\nu =0.06$~eV \citep{Esteban2020,Salas2021}, which is consistent with the 95 per cent confidence upper limit of 0.13~eV from DES~Y3.  Note that the simulations include neutrino particles using the new $\delta f$ method of \citet{Elbers2021}. 

For the purposes of the present study, it is important to highlight the inclusion of the Planck primary CMB constraints in the D3A cosmology.  As already discussed in Section \ref{sec:intro}, there is a mild tension between some LSS observables, such as cosmic shear, and the primary CMB, such that the latter prefers a larger value of $S_8$.  Hence, a joint fit to these observables will generally result in a value of $S_8$ that will be larger than preferred by cosmic shear alone, and the statistical precision of the Planck data set is such that the joint value is closer to that preferred by the primary CMB.  This will become relevant later on, when we compare the predictions of the simulatons to LSS observables.

\begin{table*}
	\centering
	\begin{threeparttable}[b]
	\caption{The values of the cosmological parameters used in different simulations. The columns list the prefix used to indicate the cosmology in the simulation name (note that for brevity the prefix `D3A' that indicates the fiducial cosmology is omitted from the simulation identifiers); the 
 dimensionless Hubble constant, $h$; the total matter density parameter, $\Omega_\text{m}$; the dark energy density parameter, $\Omega_\Lambda$; the
 baryonic matter density parameter, $\Omega_\text{b}$; the sum of the particle masses of the neutrino species, $\sum m_\nu$; the amplitude of
 the primordial matter power spectrum, $A_\text{s}$; the power-law index of the primordial matter power spectrum, $n_\text{s}$; the amplitude of the
 initial power spectrum parametrized as the r.m.s. mass density fluctuation in spheres of radius $8~h^{-1}\,\Mpc$ extrapolated to $z=0$ using linear
 theory, $\sigma_8$; the amplitude of the initial power spectrum parametrized as $S_8\equiv \sigma_8\sqrt{\Omega_\text{m}/0.3}$; the neutrino matter density 
 parameter, $\Omega_\nu \cong \sum m_\nu/(93.14~h^2\,\eV)$. Note that the values of the Hubble and density parameters are given at $z=0$. The values of the parameters that are listed in the
 last three columns have been computed from the other parameters.}
	\label{tab:cosmologies}
	\begin{tabular}{lcccccccccc} 
		\hline
		Prefix & $h$ & $\Omega_\text{m}$ & $\Omega_\Lambda$ & $\Omega_\text{b}$ & $\sum m_\nu$ & $A_\text{s}$ & $n_\text{s}$ & $\sigma_8$ & $S_8$ & $\Omega_\nu$\\
		\hline
		-            & 0.681 & 0.306 & 0.694 & 0.0486 & 0.06~eV & $2.099\times 10^{-9}$ & 0.967 & 0.807 & 0.815 & $1.39\times 10^{-3}$\\
		Planck          & 0.673 & 0.316 & 0.684 & 0.0494 & 0.06~eV & $2.101\times 10^{-9}$ & 0.966 & 0.812 & 0.833 & $1.42\times 10^{-3}$\\
		PlanckNu0p24Var & 0.662 & 0.328 & 0.672 & 0.0510 & 0.24~eV & $2.109\times 10^{-9}$ & 0.968 & 0.772 & 0.807 & $5.87\times 10^{-3}$\\
		PlanckNu0p24Fix & 0.673 & 0.316 & 0.684 & 0.0494 & 0.24~eV & $2.101\times 10^{-9}$ & 0.966 & 0.769 & 0.789 & $5.69\times 10^{-3}$\\
		LS8             & 0.682 & 0.305 & 0.695 & 0.0473 & 0.06~eV & $1.836 \times 10^{-9}$ & 0.965 & 0.760 & 0.766 & $1.39\times 10^{-3}$\\
		\hline
	\end{tabular}
	\end{threeparttable}
\end{table*}

An important aspect of our hydrodynamical simulations is the calibration of parameters which characterise the efficiencies of stellar and AGN feedback.  Following our approach in \bahamas\ \citep{McCarthy2017}, the subgrid models for BH accretion and for stellar and AGN feedback are calibrated to the observed $z=0$ galaxy stellar mass function (SMF), gas mass fractions within $R_\text{500c}$ for galaxy groups and clusters at $z\approx 0.1-0.3$ from a combination of X-ray and weak lensing data, and the $z=0$ relation between BH mass and stellar mass.  Our choice of calibration observables is motivated by the fact that the impact of baryon physics on LSS is highly correlated with the baryon fractions of galaxy groups and clusters (e.g., \citealt{Semboloni2011,Semboloni2013,Schneider2019,VanDaalen2020,Salcido2023}), as these objects dominate the matter clustering signal (e.g., \citealt{VanDaalen2015,Mead2020}).  For \flamingo\ we use a systematic Bayesian approach to the fitting that has previously been applied to the semi-analytic model GALFORM \citep[][]{Bower2010,Rodrigues2017} and to a variety of cosmological emulators based on gravity-only simulations (e.g., \citealt{Heitmann2014,Lawrence2017,Euclid2019}).  As described in \citet{Kugel2023}, we employ machine learning to fit the subgrid parameters to the calibration data. We use Gaussian process emulators trained on 32-node Latin hypercubes of simulations. The 32 nodes are distributed approximately randomly in the hypercube so that the minimum distance between the nodes is maximized. A hydrodynamical simulation is run for each node and we then build a separate emulator for each observable based on all 32 simulations. The SMF emulator takes as input the stellar mass, $M_*$, and the subgrid parameter vector, $\bmath{\theta}$, and it predicts the SMF, $f(M_*)$. The inputs for the gas fraction emulator are the total group/cluster mass, $M_\text{500c}$ (i.e.\ the mass inside the radius $R_\text{500c}$ within which the mean density is 500 times the critical density), and the subgrid parameters $\bmath{\theta}$. It outputs the gas fraction as a function of mass, $f_\text{gas,500c}(M_\text{500c})$.  

We re-calibrate the feedback model as the resolution of the simulations is varied.  This is motivated by the fact that a higher-resolution simulation resolves smaller scales and higher gas densities and will therefore, for example, yield different radiative losses and different BH accretion rates (and hence different AGN feedback), which will, in turn, change the structure of the interstellar medium even on scales resolved by the lower-resolution run (e.g., \citealt{Schaye2015}).  Another novel aspect of our approach is that the calibration takes into account the expected observational errors and biases. We impose random errors on the simulated stellar masses to account for Eddington bias. During the calibration of the fiducial intermediate-resolution model we fit for systematic errors in the SMF due to cosmic variance, bias in the inferred stellar mass, and for hydrostatic mass bias in the cluster gas fractions inferred from X-ray observations. The best-fitting bias factors, which are negligible for cosmic variance and stellar mass, and which is consistent with the literature for the hydrostatic mass bias, are then applied to the calibration data for all resolutions and models.

Note that the emulators are not only used to design simulations that reproduce the observations, but also to create models in which the SMF and/or cluster gas fractions are shifted to higher/lower values. This allows us to specify model variations in terms of the number of $\sigma$ by which they deviate from the calibration data, which is more intuitive and useful than specifying simulations solely by the values of subgrid parameters that are not directly observable. \flamingo\ includes four models in which cluster gas fractions are varied (by $+2$, $-2$, $-4$ and $-8\sigma$, respectively) while keeping the SMF unchanged, one model in which the SMF is reduced by decreasing the stellar masses by the expected systematic error (0.14~dex; \citealt{Behroozi2019}) while keeping gas fractions fixed, and two models that simultaneously vary the gas fractions and the SMF. Finally, two models use jet-like AGN feedback rather than the fiducial isotropic and thermal AGN feedback, one of which is calibrated to the fiducial data and one to gas fractions shifted down by $4\sigma$. Comparison of these last two models with the corresponding fiducial ones enables estimates of the uncertainty due to differences in the implementation of AGN feedback for a common calibration. 

\flamingo\ includes 4 intermediate-resolution hydrodynamical simulations with the fiducial calibration of the subgrid physics in $(1~\Gpc)^3$ volumes that vary the cosmological parameters. Three of the alternative cosmologies we consider are variations on \citet{Planck2020cosmopars}: their best-fitting $\Lambda$CDM model with the minimum allowed neutrino mass,  $\sum m_\nu = 0.06$~eV (`Planck'); 
a model with a high neutrino mass, $\sum m_\nu = 0.24$~eV, (allowed at 95 per cent confidence by Planck) in which the other cosmological parameters take their corresponding best-fitting values from the Planck MCMC chains (`PlanckNu0p24Var'); and a model with the same high neutrino mass, $\sum m_\nu = 0.24$~eV, that keeps all other parameters fixed to the values of model Planck, except for $\Omega_\text{CDM}$ which was reduced in order to keep $\Omega_\text{m}$ constant (`PlanckNu0p24Fix').   Note that for the latter model we fix the primordial power spectrum amplitude, $A_s$, rather than $S_8$. All models with $\sum m_\nu = 0.24$~eV use three massive neutrino species of 0.08~eV. Finally, we include the `lensing cosmology' from \citet{Amon2023} (`LS8'). This model has a lower amplitude of the power spectrum, $S_8 = 0.766$, compared with 0.815 and 0.833 for D3A and Planck, respectively. \citet{Amon2023} show that the lensing cosmology is consistent with observations of galaxy clustering from BOSS DR12 \citep{Reid2016} and galaxy-galaxy lensing from D3A \citep{Abbott2022}, HSC Y1 \citep{Aihara2018} and KiDS 1000 \citep{Kuijken2019} over a wide range of scales, $0.15 - 60~h^{-1}\,\Mpc$, if allowances are made for theoretical uncertainties associated with baryonic feedback and assembly bias.  By contrast, they show that the Planck cosmology does not fit the same data on small scales. We note that \citet{Heymans2021} showed that the LS8 model is also consistent with KiDS-1000 cosmic shear measurements. 

The simulations are initialised at $z=31$, using multi-fluid third-order Lagrangian perturbation theory (3LPT) ICs generated with the \monofonIC{} code \citep{Hahn2020,Michaux2021}.  The ICs accurately reproduce the relative growth of the individual fluid components. For \flamingo{}, a modified version of \monofonIC{} was used that implements the effects of massive neutrinos\footnote{\url{https://github.com/wullm/monofonic}}.  We use the prescriptions for 3-fluid ICs with CDM, baryons, and massive neutrinos outlined in \citet{Elbers2022a}, which builds on the 2-fluid formalism of \citet{Rampf2021} and \citet{Hahn2021}. CDM and baryon particles are set up in a two-stage process. First, the combined mass-weighted CDM + baryon fluid is initialized with single-fluid 3LPT, accounting for the presence of neutrinos. This single fluid is then split into separate components with distinct transfer functions by perturbing the masses and velocities in accordance with the first-order compensated mode. \citet{Hahn2021} showed that discreteness errors can be suppressed by perturbing particle masses rather than displacements, thereby eliminating spurious growth of the compensated mode (see also \citealt{Bird2020,Liu2023}).  The underlying Gaussian random fields were chosen from subregions of \panphasia\ to facilitate future zoom-in resimulations \citep{Jenkins2013}. To limit cosmic variance without compromising the ability to do zooms, we used partially fixed ICs \citep{Angulo2016}, setting the amplitudes of modes with $(kL)^2 < 1025$ to the mean variance, where $k$ is the wavenumber and $L$ is the side length of the simulation box.  For a more in depth description of the ICs, please see \citet{Schaye2023}.

\subsection{Projecting to cosmological observables}
\label{sec:obs}

Below we describe how the \flamingo\ data set accompanying each simulation is projected onto cosmological observables.  In the present study, we focus on observables in spherical harmonics space, namely the angular power spectrum, and reserve a configuration-space analysis for a future study.

\subsubsection{Fiducial Limber 1D analysis}

The observed 2D angular power spectrum, $C_\ell$, between two fields at a multipole moment $\ell$ can be derived employing the Limber approximation (i.e., fluctuations are only in the plane of the sky) and integrating the relevant weighted 3D power spectrum along the line of sight (e.g., \citealt{Kaiser1992,Troster2022}):
\begin{equation}
    \label{equ:cls}
    C_\ell = \int_0^{\chi(z_\mathrm{max})} \frac{W^\textrm{A}(\chi)W^\textrm{B}(\chi)}{\chi^2} \, P_\textrm{A,B}\left(\frac{\ell+\frac{1}{2}}{\chi}, z(\chi)\right) \diff\chi \ \ ,
\end{equation}
where $P(k,z)$ is the relevant 3D power (or cross-power) spectrum, $W^\textrm{A}$ and $W^\textrm{B}$ are the window functions (or weighting kernels) of the two fields, and the integral is taken over comoving distance, $\chi$, back to $\chi(z_\mathrm{max})$ where $z_{\rm max}$ is the maximum redshift, which we specify below for each of the power spectra.  

Our focus in the present study will be on the angular auto and cross-spectra between cosmic shear\footnote{Specifically E-mode shear, which we refer to here as just shear.}, CMB lensing, and the tSZ effect.  As we have already investigated the cross-spectrum between CMB lensing and the tSZ effect in \citet{Schaye2023}, we will concentrate here on the shear, CMB lensing, and tSZ effect auto-power spectra and the shear--tSZ and shear--CMB lensing cross spectra.  Taken together with the CMB lensing--tSZ effect cross in \citet{Schaye2023}, we will therefore have examined all possible auto and cross-spectra of these three observables.  Note that when examining auto- and cross-spectra that involve cosmic shear and CMB lensing but not the tSZ effect, the relevant 3D power spectrum, $P_\textrm{A,B}$, in eqn. \ref{equ:cls} is the matter power spectrum, $P_\textrm{m,m}(k,z)$, which is computed using the matter overdensity field and has units of Mpc$^3$.  For the tSZ effect auto power spectrum, $P_\textrm{A,B}$ is the 3D electron pressure power spectrum, $P_\textrm{e,e}(k,z)$, which is computed using the electron pressure field and has units of eV$^2$ cm$^{-6}$.  For the cosmic shear--tSZ effect cross, $P_\textrm{A,B}$ is the 3D matter--electron pressure cross-spectrum, $P_\textrm{m,e}(k,z)$ and has units of Mpc$^3$ eV cm$^{-3}$.  We calculate the shot noise-subtracted auto and cross power spectra of the hydrodyamical simulations following a procedure that is equivalent to that described in the appendix of \citet{Mead2020}.  

To ensure accurate computation of the observable power spectra via the Limber approximation, 3D power spectra are output on the fly with a high redshift cadence, particularly at low redshifts where non-linear and baryonic effects are most evident.   Specifically, we adopt an output frequency of $\Delta z = 0.05$ between $z=0$ and $z=3$ (60 outputs), $\Delta z = 0.25$ between $z=3$ and $z=12$ (36 outputs), $\Delta z = 0.5$ between $z=12$ and $z=20$ (16 ouputs), and $\Delta z = 1$ between $z=20$ and $z=30$ (10 outputs).  Note that the fine sampling of $\Delta z = 0.05$ below $z=3$ was deliberately chosen to match that of the background source redshift distributions of the KiDS and DES data sets.  We have tested the convergence of our calculations by using only half of the 3D power spectra (every second one) and find that resulting power spectra agree with those from our full calculation to typically better than a percent accuracy over the range of scales examined here.

In the present study, we limit our analysis to multipoles with $\ell > 100$ (corresponding to $\theta \approx 1^\degree$), for which our assumption of a flat sky, which is implicit in eqn.~\ref{equ:cls}, is highly accurate.  This multipole limit is also motivated by the fact that our 2D lightcone-based maps (described below) have been constructed assuming a flat sky and that for some tests that we examine which have a significant high-redshift contribution (e.g., CMB lensing), the simulation box size prevents us from probing very large angular scales.

In the flat-sky limit, the window function of the shear field of the $i$-th source sample, $W^{\gamma_i}(\chi)$, may be written as (e.g., \citealt{Bartelmann2001})
\begin{equation}
    \label{equ:W-gamma}
    W^{\gamma_i}(\chi) = \frac{3}{2}\left(\frac{H_0}{c}\right)^2\Omega_\mathrm{m}\frac{\chi}{a(\chi)} \int_{z}^{z_\mathrm{max}} n_i(z') \biggl[1 - \frac{\chi(z)}{\chi(z')}\biggr] \diff z' \ \ ,
\end{equation}
where $c$ is the speed of light, $a(\chi)$ is the scale factor at comoving distance $\chi$, and $n_i(z)$ is the source redshift distribution of the sample $i$, which is normalised such that its integration from $z=0$ to $z=z_{\rm max}$ is unity.  For auto and cross-spectra involving cosmic shear, $z_{\rm max}$ is set by the maximum redshift of the observed source redshift distribution.  Here we compare to KiDS 1000 and DES Y3 samples, for which $z_{\rm max} = 3$.

The window function of the CMB lensing convergence field, $W^{\kappa_{\rm CMB}}$, can be derived from eqn.~\ref{equ:W-gamma} by replacing the source redshift distribution, $n(z)$, with the Dirac delta function (i.e., a single source plane) and integrating to yield
\begin{equation}
    \label{equ:W-kappa}
    W^{\kappa_{\rm CMB}} = \frac{3}{2}\left(\frac{H_0}{c}\right)^2\Omega_\mathrm{m}\frac{\chi}{a(\chi)} \biggl(1 - \frac{\chi}{\chi_\mathrm{CMB}}\biggr) \ \ ,
\end{equation}
where $\chi_{\rm CMB}$ is the comoving distance to the surface of last-scattering, assumed to be at $z_{\rm CMB} = 1100$.  For the CMB lensing auto power spectrum, which has a very extended window in redshift space (i.e., is sensitive to fluctuations over a wide range of distances), we integrate eqn.~\ref{equ:cls} back to the initial conditions of the simulations, corresponding to $z_{\rm max} = 31$.  We have checked that contributions from higher redshifts are negligible by using the linear power spectrum beyond this maximum redshift.  Note that when comparing to observational measurements below, we use both the CMB lensing convergence ($\kappa_{\rm CMB}$) and deflection potential ($\phi$) which, in terms of their angular power spectra, are related via $\phi_\ell = 2 \kappa_{\mathrm{CMB},\ell}/[\ell(\ell+1)]$.

The window function of the tSZ effect, $W^y(\chi)$, is
\begin{equation}
    \label{equ:W-y}
    W^y(\chi) = \frac{\sigma_\mathrm{T}}{m_\mathrm{e}c^2} \frac{1}{a^2(\chi)} \ \ ,
\end{equation}
where $\sigma_\mathrm{T}$ is the Thompson scattering cross-section and $m_\mathrm{e}$ is the electron rest mass.  As for the case of the CMB lensing power spectrum, we integrate eqn.~\ref{equ:cls} for the tSZ power spectrum back to $z_{\rm max} = 31$.  However we note that, as the tSZ effect emerges from the inverse Compton scattering of CMB photons by hot free electrons, there is essentially no contribution from redshifts greater than $z=7.8$,  which corresponds to the redshift of reionisation of the simulations.  Furthermore, over the range of scales accessible to current observations, the tSZ effect power spectrum is mainly sensitive to massive, relatively nearby clusters (e.g., \citealt{Komatsu2002,Battaglia2012,McCarthy2014}) and the integrated signal is converged beyond $z\approx3$. 

\subsubsection{Map-based 2D analysis}

We describe here an alternative, map-based (2D) analysis of the simulations, which we will compare to the fiducial 1D analysis described above.  This will provide a consistency check of our results but also allows us to get a handle on the role of cosmic variance, as our lightcones are constructed for multiple observer locations.  Note that since our lightcones in most cases are restricted to a maximum redshift of $z=3$, we do not attempt to compute the CMB lensing auto power spectrum with a 2D analysis, as a non-negligible fraction of the signal comes from beyond this redshift for that statistic.

As described in the appendix of \citet{Schaye2023}, tSZ effect maps are constructed on-the-fly by accumulating the Compton $y$ values of individual particles crossing the lightcone onto \healpix maps over fixed intervals in redshift.  To construct a total (integrated) Compton $y$ map, we simply sum these maps along the line of sight back to $z=3$, which is sufficient for the tSZ power spectrum and the tSZ--cosmic shear cross-spectra that we consider.  

To construct cosmic shear and CMB lensing convergence maps, we follow the method described in \citet{McCarthy2018}, which employs the so-called Born approximation (i.e.\ light ray paths are approximated as straight lines).  In short, for each \healpix total mass map (of which there are 60 per lightcone back to $z=3$, also produced on-the-fly), we compute a projected (2D) overdensity map, $\delta(\chi,\bmath{\theta})$.  The maps are then integrated along the line of sight weighted by the window function (lensing kernel) to yield the total convergence map: 
\begin{eqnarray} 
\kappa(\bmath{\theta}) = \int_0^{\chi(z_{\rm max})} W^{\gamma_i,\ \kappa_\mathrm{CMB}}(\chi) \ \delta(\chi,\bmath{\theta}) {\rm d}\chi \ \ ,
\label{equ:kappa} 
\end{eqnarray} 
where $W^{\gamma_i, \ \kappa_\mathrm{CMB}}(\chi)$ is the window function corresponding to either the $i$-th galaxy sample (as in eqn.~\ref{equ:W-gamma}) for cosmic shear or the CMB lensing single source plane (as in eqn.~\ref{equ:W-kappa}), and $z_{\rm max} = 3$.  Note that $z_{\rm max} = 3$ is sufficient for cross-correlations between CMB lensing, cosmic shear, and the tSZ effect.
 
We use the \namaster\footnote{\url{https://namaster.readthedocs.io/en/latest/}} package \citep{Alonso2019} to compute the auto- and cross-spectra of the dimensionless scalar (spin-0) quantities $y$ and $\kappa$.  To save computational effort, the \healpix maps here have been downsampled from $N_{\rm side}=16384$ to $N_{\rm side}=4096$, corresponding to an angular resolution of $\approx 0.86$ arcmin, which is sufficient for the comparisons to observations presented below.  When computing the spectra, we initially use a multipole moment resolution (bandpower) of $\Delta \ell = 8$ but then employ a Savitzky-Golay filter of order 3 and window size of 15 to further smooth the simulated spectra.  Note that the smoothing is applied for visual (plotting) purposes only. Unlike for the fiducial Limber 1D analysis, we do not quantitatively compare the smoothed 2D map-based power spectra to the observed power spectra (e.g., compute a goodness of fit).  We deconvolve the $N_{\rm side}=4096$ pixel window function from the computed cross-spectra using the \textit{pixwin} function within the \healpix package.  

We note that within the flat-sky limit adopted here, the $C_\ell$'s for (E-mode) shear are the same as those for the convergence, $\kappa$ (e.g., \citealt{Kilbinger2017,Wei2018}).  Thus, for our 2D cosmic shear analysis, there is no need to convert the convergence field, $\kappa(\bmath{\theta})$, in eqn.~\ref{equ:kappa} into a shear field, $\gamma_i(\bmath{\theta})$, before computing the shear power spectra\footnote{We have verified this to very high precision by converting the spin-0 convergence field into a spin-2 shear field using the method of \citet{Kaiser1993} and then computing the shear power spectra using \namaster.}.

\section{Results}
\label{sec:results}

In this section we present the main results of our analyses.  We begin by examining the power spectra of cosmic shear, tSZ effect, and CMB lensing, before examining the cross-spectra between these observables.  For each case we make use of the full \flamingo\ suite of variations, exploring the dependence of the signals on cosmology, the efficiency and nature of feedback, and simulation volume and resolution.  We compare the simulations with the most recent observational measurements of these quantities.

\begin{figure}
    \includegraphics[width=\columnwidth]{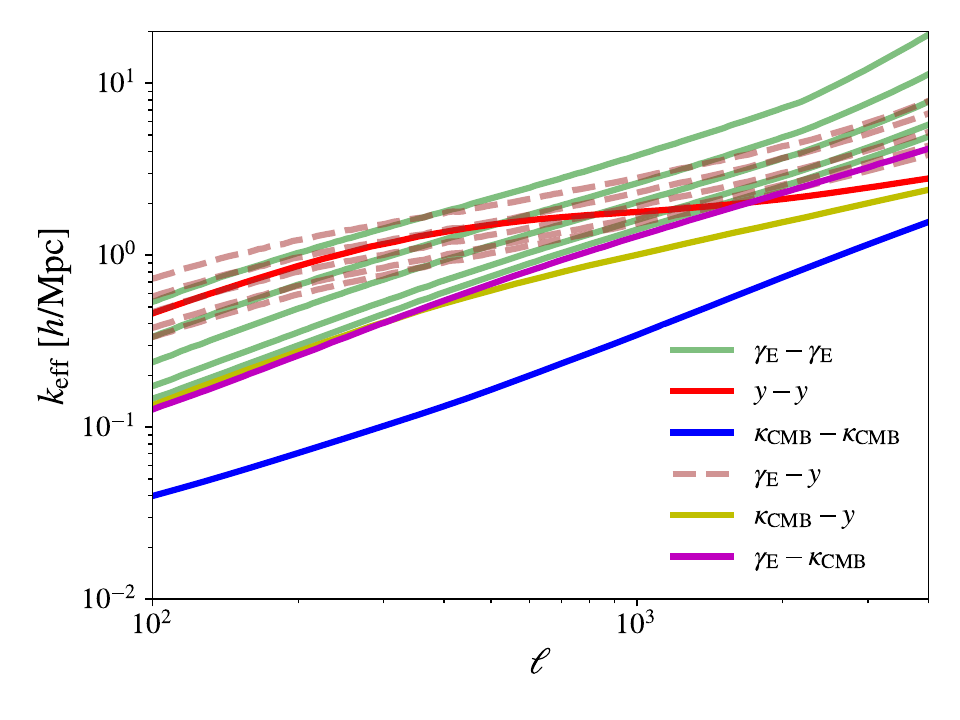}\\
    \includegraphics[width=\columnwidth]{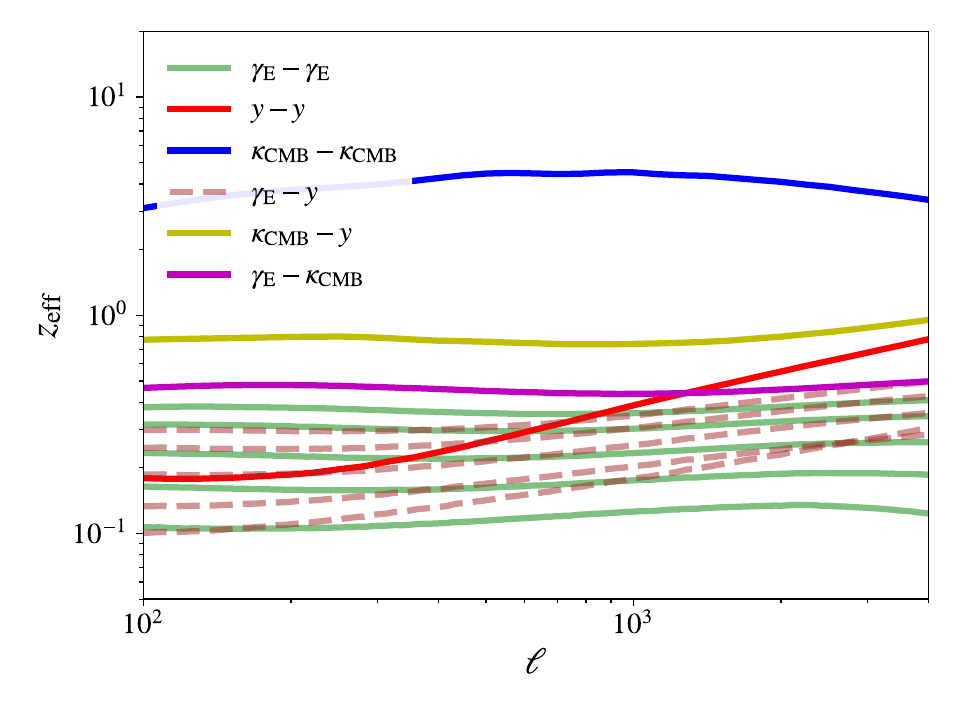}
    \caption{$C_\ell$-weighted $k$-scale ($k_{\rm eff}$; top panel) and redshift ($z_{\rm eff}$; bottom panel) of the auto and cross-power spectra involving cosmic shear, the tSZ effect, and CMB lensing considered in this paper.  For cosmic shear power spectra ($\gamma_\mathrm{E}$-$\gamma_\mathrm{E}$) we plot only the auto spectra for the 5 tomographic bins of the KiDS 1000 survey (i.e., 1-1, 2-2, ..., 5-5, shown in light green curves), where the 1-1 power spectrum has the smallest effective redshift and the largest effective $k$.  Similarly, for the cosmic shear--tSZ effect cross-spectrum (dashed brown curves), the 1-$y$ cross has the smallest $z_{\rm eff}$ and the largest $k_{\rm eff}$.  The top panel shows that $k_{\rm eff}$ rises with increasing $\ell$, though the gradient is significantly shallower for spectra involving tSZ $y$ relative to pure lensing power spectra.  In general, the cosmic shear power spectrum, tSZ effect power spectrum, and their cross probe the smallest scales (largest $k_{\rm eff}$'s) at a given $\ell$, with the KiDS 2D cosmic shear--CMB lensing cross and CMB lensing--$y$ cross probing larger scales, and the CMB lensing power spectrum probing the largest scales.  The effective redshift of the auto and cross spectra are rank ordered in the opposite sense to $k$, with the cosmic shear power spectrum, tSZ effect power spectrum, and their cross-spectrum being sensitive to the lowest redshifts (generally $z_{\rm eff} \approx 0.1-0.4$, apart from the tSZ effect power spectrum on small angular scales), followed by the KiDS 2D cosmic shear--CMB lensing cross ($z_{\rm eff} \approx 0.5$), the CMB lensing--$y$ cross ($z_{\rm eff} \approx 0.8$), and the CMB lensing power spectrum sampling the highest redshifts ($z_{\rm eff} \approx 3-4$).}
    \label{fig:keff_zeff}
\end{figure}

\subsection{Effective length scales and redshifts probed by different observables}

To help aid the interpretation of the results presented below and to assess consistency between the different tests, we show in Fig.~\ref{fig:keff_zeff} the $C_\ell$-weighted mean $k$-scale ($k_{\rm eff}$; top panel) and redshift ($z_{\rm eff}$; bottom panel) of the various auto and cross-power spectra considered here.  To compute $k_{\rm eff}$ and $z_{\rm eff}$ we use the fiducial 1D Limber integration (eqn.~\ref{equ:cls}) with 3D power spectra from the largest \flamingo\ hydro run (L2p8\_m9).  More specifically, to compute $k_{\rm eff}$ we modify eqn.~\ref{equ:cls} to include an additional multiplicative term, $k\equiv(\ell+1/2)/\chi$, and then integrate the modified equation over $\chi$.  We then divide the result by the integration of the unmodified eqn.~\ref{equ:cls}, to yield the $C_\ell$-weighted $k$ scale, $k_{\rm eff}$.  The calculation of $z_{\rm eff}$ is performed in an analogous way, but using $z(\chi)$ as the additional multiplicative term.
Note that for cosmic shear ($\gamma_\textrm{E}-\gamma_\textrm{E}$), we use source redshift distributions of the 5 tomographic bins of the KiDS 1000 survey (\citealt{Troster2022}; as discussed below in Section \ref{sec:shear-ps}), apart from the shear--CMB lensing cross, for which we use the combined (single bin) distribution from \citet{Robertson2021}, as described in Section \ref{sec:shear-CMB-ps}.  

The top panel of Fig.~\ref{fig:keff_zeff} shows that $k_{\rm eff}$ generally rises with increasing $\ell$ as expected, though the gradient is significantly more shallow for spectra involving tSZ $y$ relative to pure lensing power spectra.  In general, the cosmic shear power spectrum, tSZ effect power spectrum, and their cross-spectrum probe the smallest scales (largest $k_{\rm eff}$'s) at a given multipole, followed by the KiDS 2D cosmic shear--CMB lensing cross and CMB lensing--$y$ cross which probe larger scales, and the CMB lensing power spectrum which samples the largest scales at a given multipole.  

The bottom panel of Fig.~\ref{fig:keff_zeff} shows the effective redshifts of the auto and cross spectra are rank ordered in the opposite sense to $k$, with the cosmic shear power spectrum, tSZ effect power spectrum, and their cross-spectrum being sensitive to the lowest redshifts (generally $z_{\rm eff} \approx 0.1-0.4$, apart from the tSZ effect power spectrum on small angular scales which probes somewhat larger redshifts), followed by the KiDS 2D cosmic shear--CMB lensing cross ($z_{\rm eff} \approx 0.5$), the CMB lensing--$y$ cross ($z_{\rm eff} \approx 0.8$) explored in \citet{Schaye2023}, and the CMB lensing power spectrum being sensitive to the highest redshifts ($z_{\rm eff} \approx 3-4$).  Thus, a simultaneous analysis of all these statistics will sample fluctuations over very wide ranges of redshift and physical scale, and in this respect there is a significant degree of overlap but also complementarity between the different observables.  

Another relevant dimension which we do not consider in Fig.~\ref{fig:keff_zeff}, but which would be fruitful to examine in future studies in the context of the halo model, is the halo mass and radial dependence of the various auto- and cross-spectra.  For example, while the cosmic shear and tSZ effect power spectra may probe similar physical scales and redshifts, they are known to depend quite differently on halo mass and this is likely to be important in the context of potential baryon/feedback effects.  When describing the impact of baryons on the various power spectra below, we will provide a qualitative link to the role of halo mass, leaving a quantitative exploration for future work.

\begin{figure*}
    \includegraphics[clip, trim=3.0cm 2.5cm 3.0cm 4.0cm, width=\textwidth]{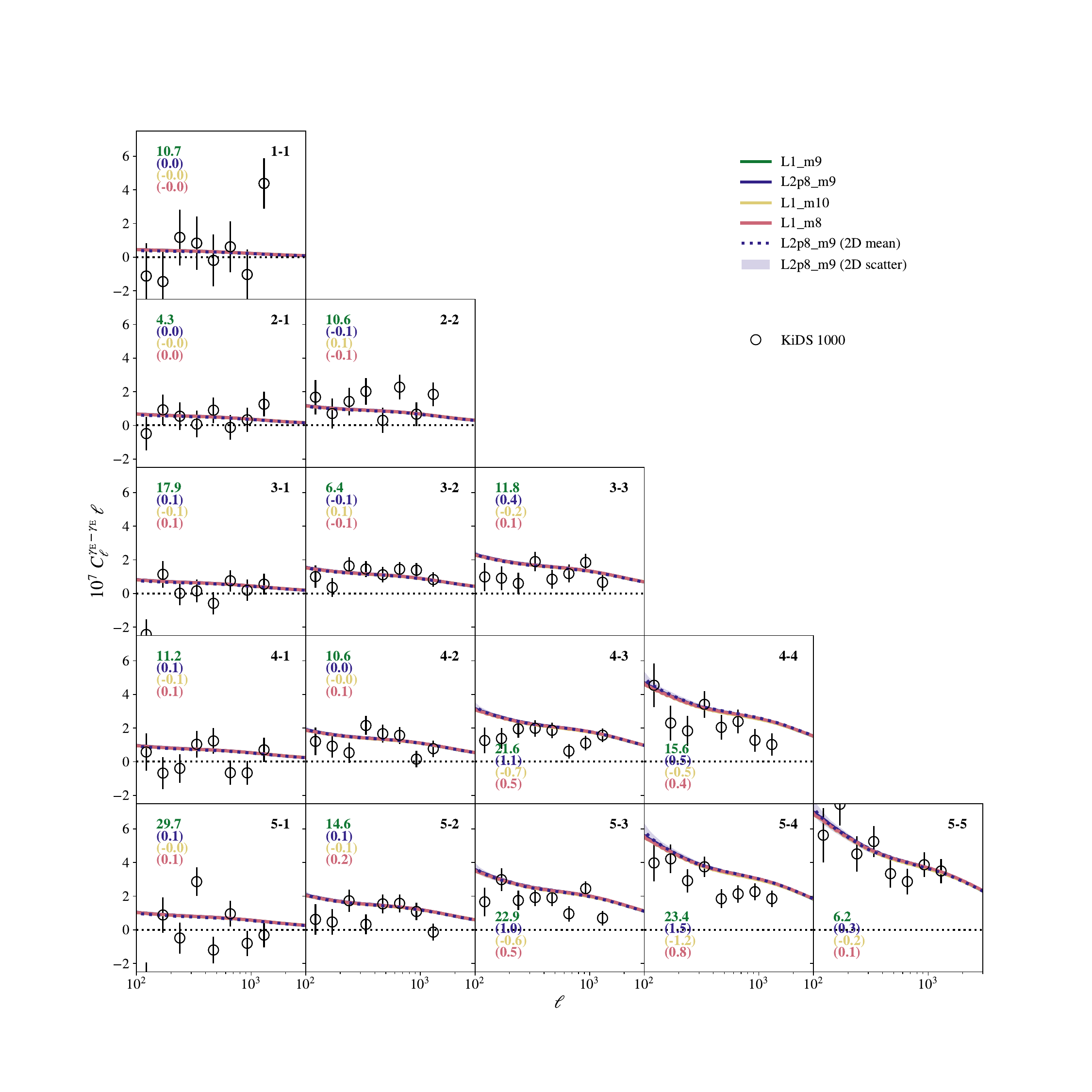}
    \caption{ Dependence of the predicted KiDS 1000 cosmic shear power spectrum on simulation box size and resolution.  The numbers in the top right corner of each panel indicate the tomographic bins being used (e.g., 3-2 indicates a cross-spectrum between the third and second tomographic bins, which is equivalent to 2-3).  The open circles correspond to the KiDS 1000 measurements of \citet{Troster2022} and the error bars correspond to the diagonal components of the covariance matrix.  The solid coloured curves correspond to the predicted spectra for the \flamingo\ simulations with the fiducial D3A cosmology as the box size and resolution are varied.  Following the same colour coding, the numbers on the left of each panel indicate $\chi^2$ for the L1\_m9 run and the $\Delta \chi^2$ (in parentheses) of the other runs with respect to L1\_m9.  Note that the $\chi^2$'s are computed here using the diagonal elements of the covariance matrix and we compute them for the fiducial 1D Limber method only (solid curves).  A negative value for the $\Delta \chi^2$ indicates a better match to the data.  The blue dashed curve and shaded region correspond to the mean power spectra and scatter from the 2D map-based analysis of the 8 light cones for the L2p8\_m9 run, which has the same resolution, cosmology, and calibrated feedback model as L1\_m9.    The simulation predictions are converged with box size and resolution, but tend to predict spectra with slightly higher amplitude than observed (i.e., the $S_8$ tension), particularly for the spectra involving the higher tomographic bins.}
    \label{fig:shear_powerspec_1}
\end{figure*}

\subsection{Cosmic shear power spectrum}
\label{sec:shear-ps}

We begin by comparing the \flamingo\ simulations to measurements of the cosmic shear power spectrum from the KiDS 1000 survey\footnote{\url{https://kids.strw.leidenuniv.nl/}} \citep{Kuijken2019,Heymans2021} from the recent study of \citet{Troster2022}.  The background source galaxies are divided into 5 tomographic bins based on their photometric redshifts, which are derived from nine-band imaging data spanning optical to infrared wavelengths \citep{Hildebrandt2021}.  The tomographic bins, labelled 1 through 5, have selection windows of $0.1 < z_{\rm B} < 0.3$, $0.3 < z_{\rm B} < 0.5$, $0.5 < z_{\rm B} < 0.7$, $0.7 < z_{\rm B} < 0.9$, and $0.9 < z_{\rm B} < 1.2$, respectively, where $z_{\rm B}$ corresponds to the maximum in the redshift posterior probability distribution for individual galaxies.  The source redshift distributions, $n_i(z)$, for the 5 tomographic bins are shown in figure 1 of \citet{Troster2022}.  We use these distributions when computing the shear window function (eqn.~\ref{equ:W-gamma}) in order to project the simulation 3D power spectra (or 2D maps) onto shear power spectra for comparison to the KiDS 1000 measurements.  \citet{Troster2022} measure the auto- and cross-power spectra between the 5 tomographic bins.  Following their analysis, we adopt angular scale cuts of $100 < \ell < 1500$, corresponding to the range over which the KiDS cosmic shear methodology has been validated.  \citet{Troster2022} have made the measurements, covariance matrices, source redshift distributions, and their analysis software publicly available\footnote{\url{https://github.com/tilmantroester/KiDS-1000xtSZ}}.  These authors have also measured the shear--tSZ effect cross-spectrum, which we compare to in Section \ref{sec:shear-tSZ-ps} below.

\begin{figure*}
    \includegraphics[clip, trim=3.0cm 2.5cm 3.0cm 4.0cm, width=\textwidth]{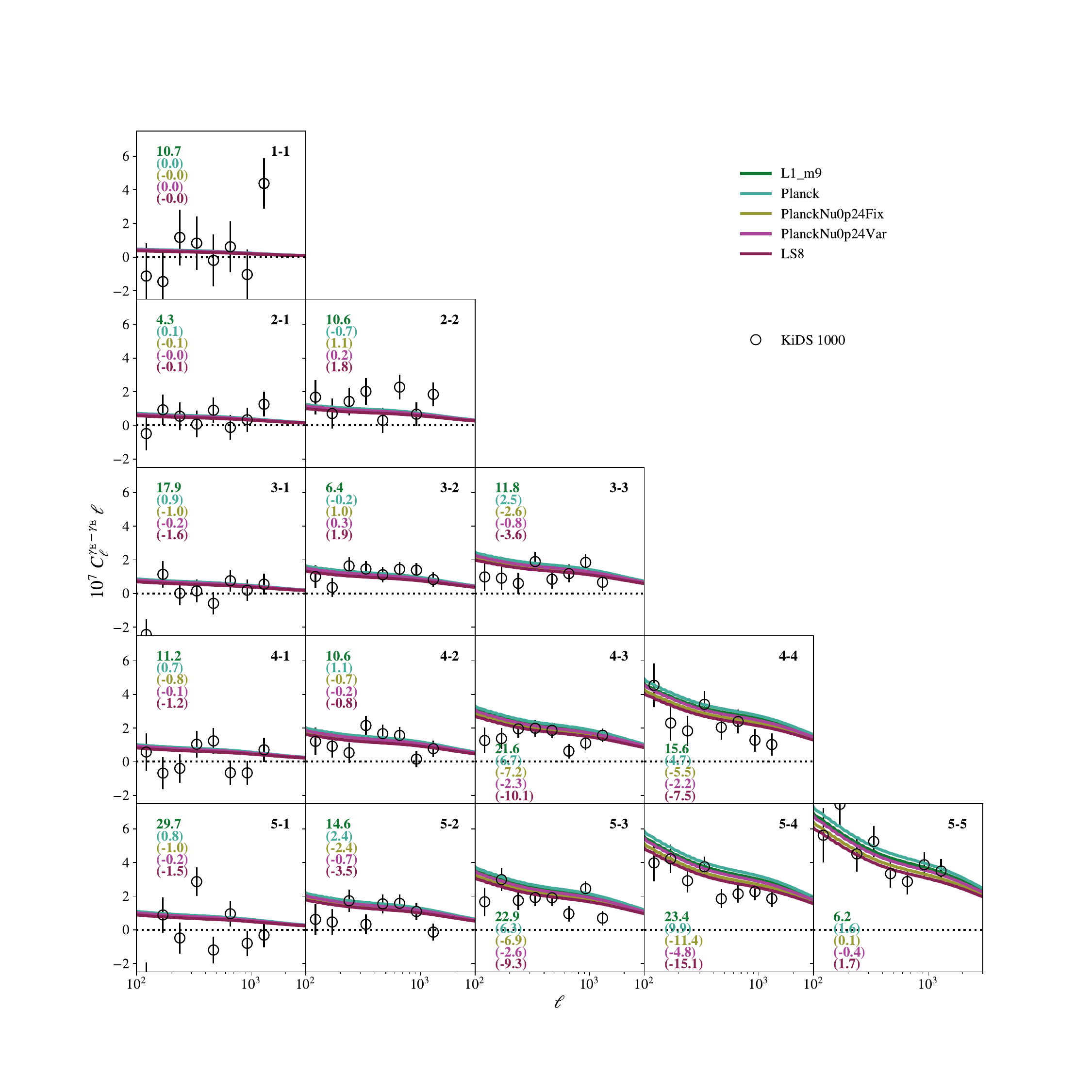}
    \caption{As Fig.~\ref{fig:shear_powerspec_1}, but showing the dependence of the predicted KiDS 1000 cosmic shear power spectrum on cosmology.  The solid coloured curves correspond to the predicted spectra for the \flamingo\ simulations as the background cosmology is varied.  A Planck maximum-likelihood cosmology yields spectra with elevated power relative to the fiducial D3A (and is thus in stronger tension with the lensing measurements), whereas increasing the summed neutrino mass from 0.06 eV (fiducial) to 0.24 eV suppresses the power over all scales sampled here.  The LS8 `lensing' cosmology predicts the lowest power and is in best agreement with the measurements, particularly for the high tomographic bin data (except 5-5).}
    \label{fig:shear_powerspec_2}
\end{figure*}

In Fig.~\ref{fig:shear_powerspec_1} we examine the box size and resolution dependence of the predicted cosmic shear power spectra.  The numbers in the top right corner of each panel indicate the tomographic bins being used (e.g., 3-2 indicates a cross-spectrum between the third and second tomographic bins, which is equivalent to 2-3).  The open circles correspond to the KiDS 1000 measurements of \citet{Troster2022} and the error bars shown correspond to the diagonal components only of the covariance matrix.  The solid coloured curves correspond to the predicted spectra for the \flamingo\ calibrated hydro simulations with the fiducial D3A cosmology while the box size and resolution are varied.  Following the same colour coding, the numbers on the left of each panel indicate $\chi^2$ for the L1\_m9 run and the $\Delta \chi^2$ (in parentheses) of the other runs with respect to L1\_m9.  Note that the $\chi^2$'s are computed here adopting the diagonal elements of the covariance matrix and we compute them for the fiducial 1D Limber method only (solid curves).  The blue dashed curve and shaded region correspond to the mean power spectra and scatter from the 2D map-based analysis of the 8 light cones for the L2p8\_m9 run, which has the same resolution, cosmology, and calibrated feedback model as L1\_m9.  Generally speaking, the predicted signal is strongest for spectra involving the higher tomographic bin numbers, simply as a result of there being a longer path length along the line of sight from the observer to the galaxy samples and thus more lensing.

From Fig.~\ref{fig:shear_powerspec_1} we conclude that the predictions of the simulations are robust to variations in box size and resolution, as well analysis method (1D Limber vs.~2D map-based).  There is perhaps a hint that the larger 2.8 Gpc volume has slightly more power on the largest scales ($\ell \la 500$) for the spectra involving the most distant tomographic bins.  However, these differences are generally small compared to variations of the simulations with respect to the KiDS 1000 measurements, variations between neighbouring data points, and the estimated uncertanties.

Also evident from Fig.~\ref{fig:shear_powerspec_1} is the tendency of the simulations to predict power spectra that are elevated with respect to the KiDS 1000 measurements, particularly for the higher tomographic bins from which much of the signal originates.  This is confirmation of the well-known $S_8$ tension, noting again that the D3A cosmology includes Planck CMB constraints which pull the preferred value of $S_8$ to higher values than favoured by cosmic shear alone.  However, we highlight here that our conclusions are only qualitative, as we have not taken into account the uncertainties in the D3A cosmology, nor have we marginalised over relevant systematic uncertainties in the lensing measurements (e.g., intrinsic alignments, photo-$z$ uncertainties, etc.).

In Fig.~\ref{fig:shear_powerspec_2} we examine the \flamingo\ cosmological variations in the context of the fiducial calibrated hydro model.  Specifically, in 1 Gpc volumes we compare the fiducial D3A cosmology with the maximum likelihood Planck 2018 cosmology (`Planck'), two Planck-based cosmologies where the summed mass of neutrinos is raised from the minimum value of 0.06 eV to 0.24 eV (see Table \ref{tab:cosmologies}), and the `lensing' cosmology (LS8) of \citet{Amon2023}, which uses CMB data to inform the mass densities of baryons and CDM as well as the primordial power spectrum shape ($n_s$), but uses cosmic shear to set the amplitude ($S_8$, or $A_s$).

The dependence on cosmology is particularly evident in the spectra involving the higher tomographic bins, with the Planck cosmology yielding spectra that are elevated with respect to the fiducial D3A cosmology (and thus in slightly stronger tension with the KiDS 1000 measurements), whereas increasing the summed neutrino mass suppresses the power over all scales sampled here.  Note that increasing the neutrino mass and lowering $S_8$ have similar effects on the cosmic shear power spectra.  This is a consequence of our approach of fixing $A_s$ (i.e., using the CMB to specify it) when increasing the summed neutrino mass, as the main effect of increasing the neutrino mass in this case is to suppress the clustering amplitude at late times (i.e., lower $S_8$).  The LS8 `lensing' cosmology, with $S_8=0.766$, predicts the lowest power and is in best agreement with the measurements, which is essentially by construction.  Our findings appear consistent with those of \citet{Troster2022}, who obtain $S_8 \approx 0.75 \pm 0.02$ using the \textsc{HMx} model of \citet{Mead2020} to model the cosmic shear power spectrum and cosmic shear--tSZ cross-spectrum.  Note that \textsc{HMx} includes a model for marginalising over the impact of baryons which was calibrated on our previous \bahamas\ simulations.

\begin{figure*}
    \includegraphics[clip, trim=3.0cm 2.5cm 3.0cm 4.0cm, width=\textwidth]{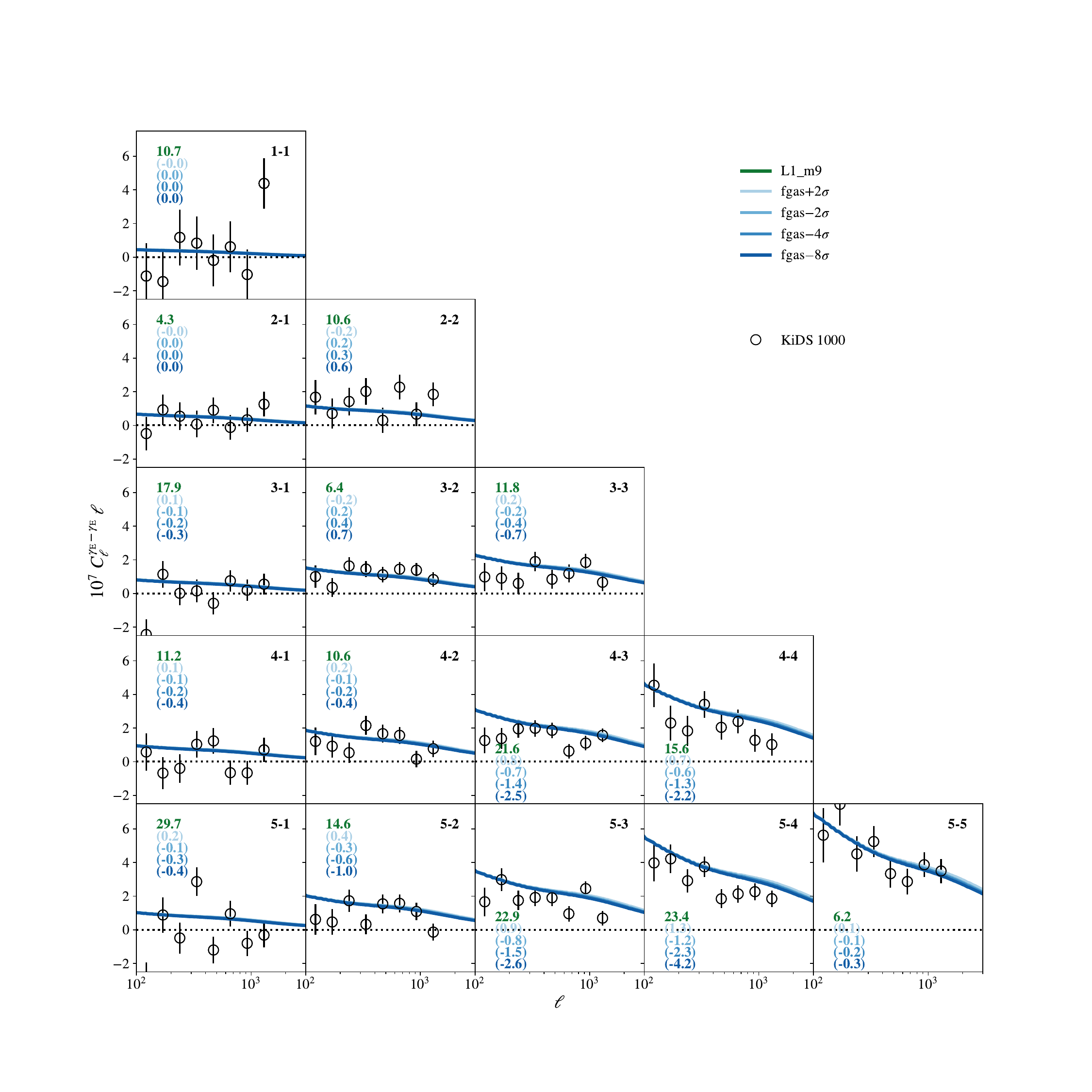}
    \caption{As Fig.~\ref{fig:shear_powerspec_1}, but showing the dependence of the predicted KiDS 1000 cosmic shear power spectrum on baryon physics, namely variations in the gas fractions of groups and clusters which are mediated primarily through variations in the AGN feedback strength.  The solid coloured curves correspond to the predicted spectra for the \flamingo\ simulations as the gas fractions are varied from $+2\sigma$ to $-8\sigma$ with respect to the observed gas fraction--halo mass relation (see \citealt{Schaye2023,Kugel2023}).  Lowering the gas fractions (increasing the feedback strength) relative to the fiducial calibrated model results in a slightly improved match to the data, though the improvement is generally small compared to that from the investigated changes in the baseline cosmology (Fig.~\ref{fig:shear_powerspec_2}).}
    \label{fig:shear_powerspec_3}
\end{figure*}

We now turn to the impact of variations in baryon physics on the cosmic shear power spectrum.  In Fig.~\ref{fig:shear_powerspec_3} we show how varying the gas fractions of groups and clusters (as mediated through variations in stellar and primarily AGN feedback) affects cosmic shear.  Even though the feedback variations span a wide range ($+2\sigma$ to $-8\sigma$ about the observed gas mass fraction--halo mass relation; see \citealt{Schaye2023,Kugel2023}), they result in only relatively minor effects on the power spectra, which are most evident at $\ell \ \ga \ 700$.  In particular, reducing the gas fractions slightly improves the match to the measurements of the high tomographic bin data.  Note, however, that the improvement is generally small compared to that which is obtained from variations in the baseline cosmology.  Indeed, the slight preference for increased feedback with respect to the fiducial calibrated model is most likely driven by the fact that there is an offset with respect to the baseline D3A cosmology generally (that is, increased feedback is partially compensating for a difference in cosmology).  We can test this by assuming the impact of cosmology variations is separable from the impact of baryon variations \citep{VanDaalen2011,Mummery2017,VanDaalen2020}.  Specifically, we compute the `suppression function' of the fgas$-8\sigma$ run with respect to the fiducial hydro model (both in the fiducial D3A cosmology) by simply taking the ratio of their power spectra (e.g., \citealt{Semboloni2011}).  Assuming this ratio is independent of cosmology, we multiply it with the LS8 power spectrum (which uses the fiducial calibrated hydro model).  This procedure approximates the impact of running the LS8 cosmology but with stronger feedback.  Comparing the unmodified LS8 and its enhanced feedback variant to the observational measurements, we find that the preference for stronger feedback largely goes away.  For example, in the 5-3 and 5-4 cross spectra cases, which have the strongest preference for increased feedback in the fiducial D3A cosmology, the $\Delta \chi^2$'s between the LS8 model and its increased feedback (fgas$-8\sigma$) variant are only $-0.9$ and $-1.5$, compared to $-2.6$ and $-4.2$ in the D3A cosmology in Fig.~\ref{fig:shear_powerspec_3}.  Thus, we confirm that cosmology and baryon feedback can be degenerate when fitting to the observational measurements, which underscores the importance of external data sets (e.g., group baryon fractions) in constraining baryonic feedback.

We examine other baryonic feedback scenarios in Fig.~\ref{fig:shear_powerspec_4} in Appendix \ref{sec:other}, namely the impact of variations in the SMF (both at fiducial and reduced gas fractions) and the calibrated and stronger jet model of AGN feedback.  Similar to what was concluded above, we find that these variations in the feedback models have generally a minor impact on the cosmic shear power spectrum.  Variations in the SMF at fixed gas fraction have a negligible impact on all scales, which is not unexpected as the baryon fractions of groups and clusters are generally dominated by the hot gas.  The jet models yield similar results to the fiducial thermal AGN model at fixed gas fraction.

\begin{table}
	\centering
	\caption{$\chi^2$ values of the predicted KiDS 1000 and DES Y3 cosmic shear power spectra with respect to the measurements of \citet{Troster2022} and \citet{Doux2022}, respectively.  There are 120 (270) independent data points summed over the 15 (10) KiDS 1000 (DES Y3) auto- and cross-spectra.  Values in parentheses indicate the $\Delta \chi^2$ with respect to the L1\_m9 run.  $\chi^2$ values using either the diagonal elements alone or the full covariance matrix are provided.}
	\label{tab:shear_chi2}
	\begin{tabular}{lrrrr} 
        & \multicolumn{2}{c}{KiDS 1000} & \multicolumn{2}{c}{DES Y3}\\ 
        \hline
		Prefix & $\chi^2_{\rm diag}$ & $\chi^2_{\rm covar}$ & $\chi^2_{\rm diag}$ & $\chi^2_{\rm covar}$\\
		\hline
            L1$\_$m9 & 217.5 & 183.2 & 355.2 & 324.3\\
            L2p8$\_$m9 & (5.1) & (2.2) & (11.3) & (3.5)\\
            L1$\_$m10 & (-3.7) & (-1.7) & (-6.5) & (-2.8)\\
            L1$\_$m8 & (2.8) & (1.4) & (3.9) & (1.2)\\
            fgas$+2\sigma$ & (4.7) & (2.4) & (5.6) & (1.3)\\
            fgas$-2\sigma$ & (-4.0) & (-2.0) & (-4.5) & (-1.0)\\
            fgas$-4\sigma$ & (-7.8) & (-3.8) & (-8.4) & (-1.7)\\
            fgas$-8\sigma$ & (-13.7) & (-6.7) & (-14.0) & (-2.6) \\
            M*$-\sigma$ & (-2.6) & (-1.3) & (-2.8) & (-0.5) \\
            M*$-\sigma$\_fgas$-4\sigma$ & (-9.3) & (-4.6) & (-9.6) & (-1.7)\\
            Jet & (0.8) & (0.6) & (-0.5) & (-0.7)\\
            Jet\_fgas$-4\sigma$ & (-14.8) & (-7.2) & (-16.9) & (-4.2)\\
            Planck & (36.9) & (16.1) & (76.4) & (24.2)\\
            PlanckNu0p24Fix & (-37.3) & (-16.3) & (-43.9) & (-13.1)\\
            PlanckNu0p24Var & (-14.2) & (-6.3) & (-21.7) & (-7.5)\\
            LS8 & (-48.8) & (-21.2) & (-34.4) & (-8.5) \\
            \hline
	\end{tabular}
\end{table}

The DES Y3 release provides an independent data set with similar statistical precision to the KiDS 1000 survey against which we can compare the simulations to test for consistency (or lack thereof).  In Appendix \ref{sec:other} we compare the \flamingo\ simulations with the DES Y3 harmonic space measurements of \citet{Doux2022}.  As shown in Fig.~\ref{fig:shear_powerspec_DES}, the lensing LS8 cosmology yields a somewhat better fit to the data relative to the fiducial D3A cosmology (particularly amongst the majority of the higher tomographic bins), whereas a Planck CMB cosmology yields a worse fit for virtually all bins.  Also consistent with the KiDS 1000 comparison above, increasing the efficiency of feedback slightly improves the fit to the DES measurements, but is less significant than the improvement that results from lowering of $S_8$.  We conclude that while there is evidence for (mild) tension of the DES Y3 measurements with the Planck CMB cosmology, it is of slightly lower significance than for the KiDS 1000 survey.  This is consistent with findings of \citet{Doux2022}, who infer $S_8 = 0.784 \pm 0.026$ from the DES Y3 power spectra, representing a $1.5\sigma$ tension with Planck.  We note that KiDS 1000 and DES Y3 shear results are consistent with each other to within $1\sigma$ (e.g., \citealt{Abbott2023}).

For completeness, in Table \ref{tab:shear_chi2} we present the $\chi^2$ values of the predicted cosmic shear power spectra with respect to the observational KiDS 1000 and DES Y3 measurements.  The values are summed over the 15 and 10 auto- and cross-spectra from KiDS and DES, respectively, and are shown for both the cases where only the diagonal errors are adopted and when the full covariance matrix is used.  The results confirm those shown in the figures, although the differentiability of the models is reduced when taking into account the full covariance of the measurements (as expected).

\subsection{tSZ effect power spectrum}
\label{sec:tSZ-ps}

We now turn our attention to the tSZ effect power spectrum.  The tSZ effect is induced from the inverse Compton scattering of CMB photons by hot, free electrons in the intracluster medium (ICM) of galaxy groups and clusters \citep{Sunyaev1972,Birkinshaw1999}.  The effect appears as a decrement in CMB temperature maps at radio wavelengths and an increment at millimetre scales (for a review see \citealt{Carlstrom2002}).
As its amplitude is proportional to both the ICM electron number density and temperature, it is particularly strong for massive galaxy clusters.  Indeed, the self-similar expectation is that the integrated tSZ flux scales with halo mass to the $5/3$ power (e.g., \citealt{White2002,Planck2013}) and previous studies have shown that massive clusters tend to dominate the tSZ effect power spectrum (e.g., \citealt{Komatsu2002,Battaglia2012,McCarthy2014}).  As the abundance of massive clusters is a sensitive probe of $\Omega_m$ and particularly $\sigma_8$, the tSZ effect, which is proportional to the square of the tSZ flux, is even more sensitive to these cosmological parameters (e.g., \citealt{Komatsu2002,Shaw2010,Millea2012}).  For example, the amplitude of the tSZ effect power spectrum scales approximately as $\sigma_8^{8.3}$ \citep{Shaw2010}.

We compare the \flamingo\ simulations to the latest tSZ effect power spectrum measurements, namely the Planck-based measurements reported in \citet{Bolliet2018} and the South Pole Telescope (SPT) data in \citet{Reichardt2021}.  Note that \citet{Bolliet2018} present an improved re-analysis of the Planck 2015 tSZ data set from \citet{Planck2016_tSZ_PS}, by taking into account the tri-spectrum in the covariance matrix and placing physical constraints on the amplitudes of foreground contaminants (particularly radio and infrared point sources and the clustered infrared background, or CIB).  We use the tabulated tSZ power spectrum measurements and total diagonal uncertainties from table 4 of \citet{Bolliet2018}.  Note that total uncertainties include the non-Gaussian contribution from the tri-spectrum which dominates on large scales \citep{Komatsu2002}.  The full covariance matrix was not tabulated, though as shown in figure 3 of \citet{Bolliet2018} the diagonal uncertainties tend to dominate.

\begin{figure*}
    \includegraphics[width=\columnwidth]{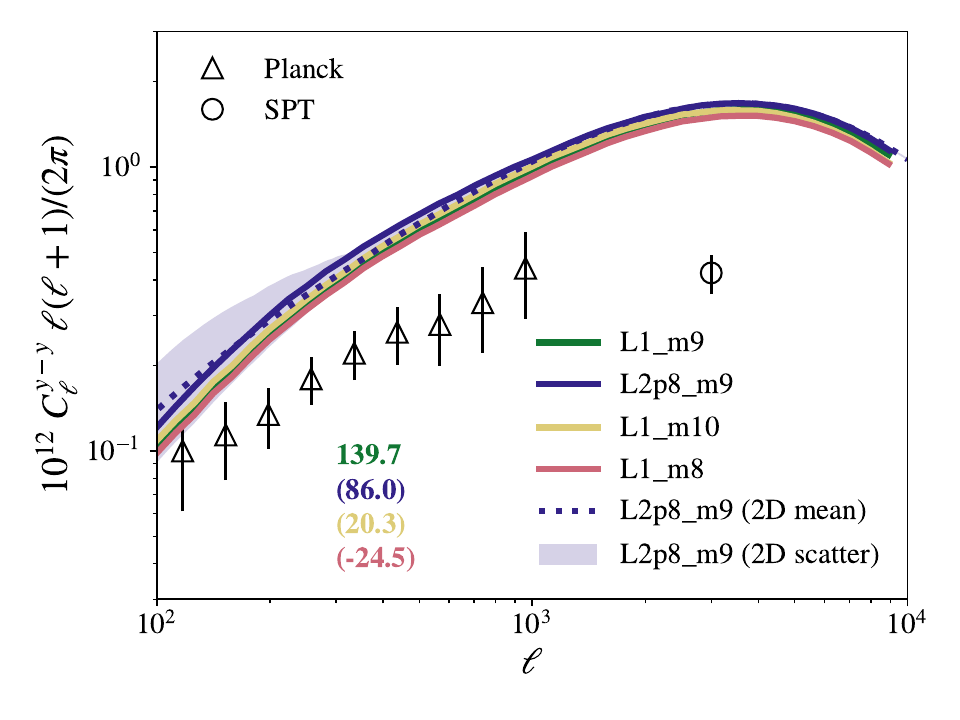}
    \includegraphics[width=\columnwidth]{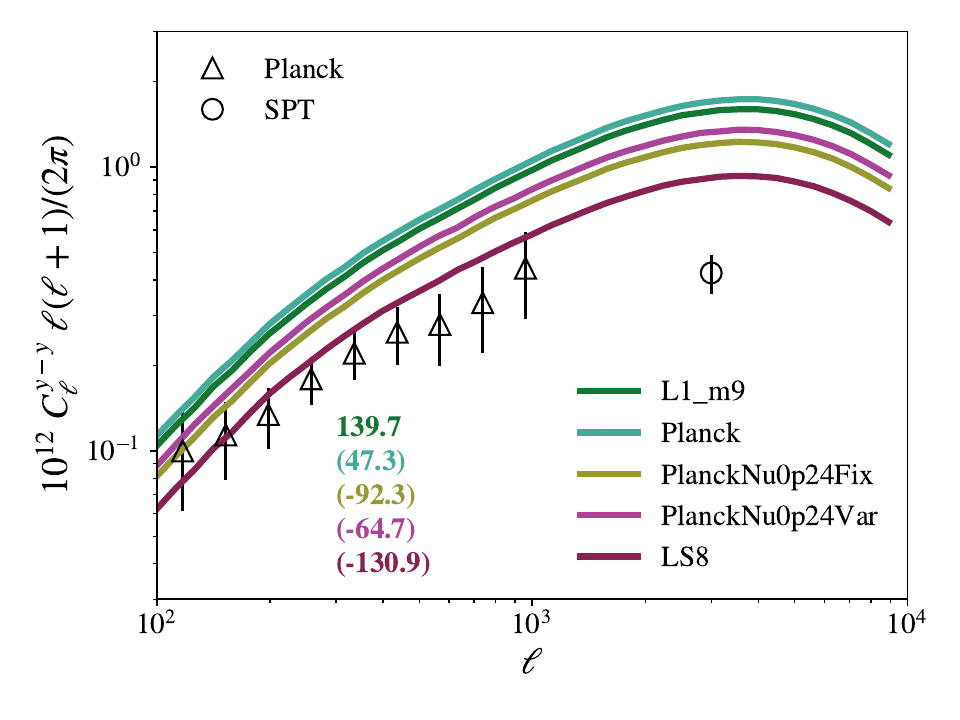}\\
    \includegraphics[width=\columnwidth]{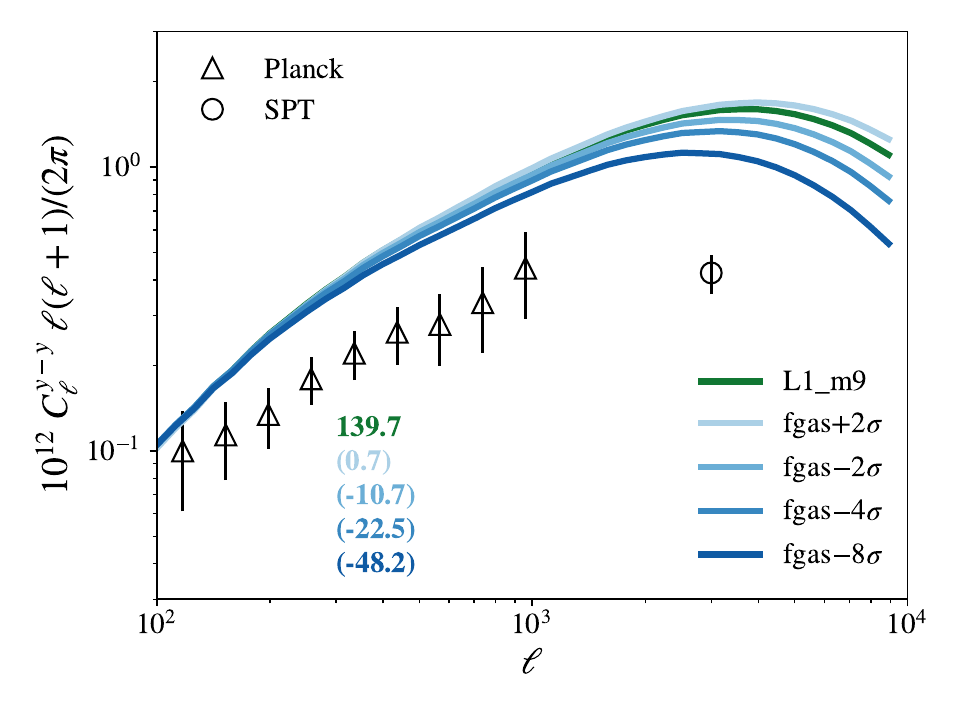}
    \includegraphics[width=\columnwidth]{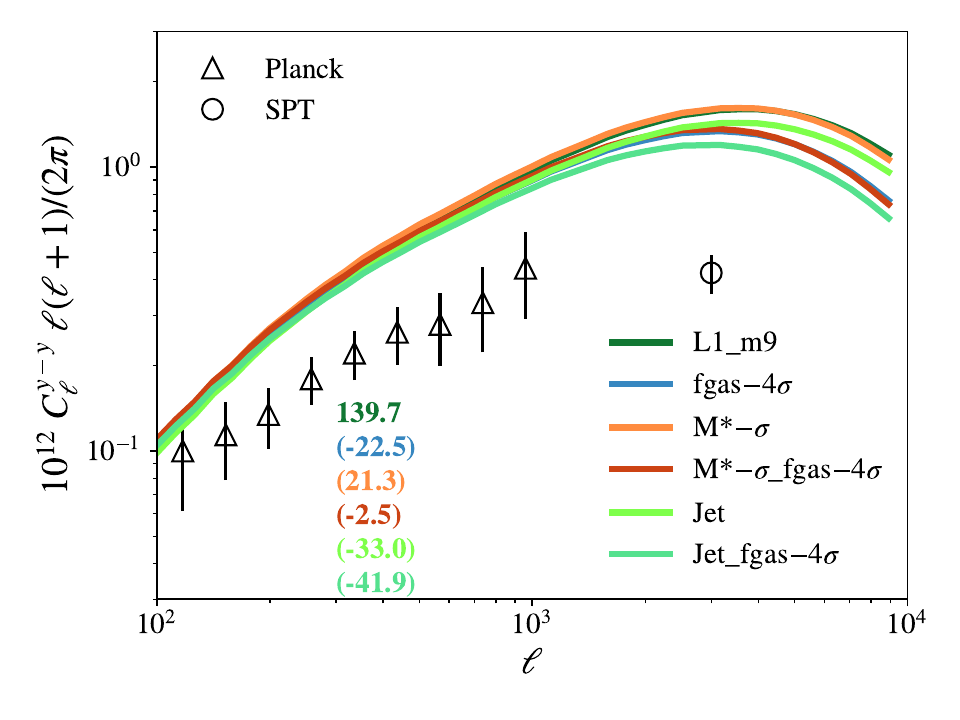}
    \caption{The tSZ effect angular power spectrum.  The open triangles correspond to the Planck tSZ measurements of \citet{Bolliet2018} while the open circle corresponds to the SPT measurements of \citet{Reichardt2021}.  \textit{Top left:} Dependence on simulation box size and resolution.   Note that the $\chi^2$ values displayed in the bottom left are computed with respect to the Planck tSZ measurements are take into account only the diagonal elements of the covariance matrix.  The simulation predictions are largely converged with resolution, though the 1 Gpc boxes are missing a small amount of power compared to the 2.8 Gpc volume. \textit{Top right:} Dependence on cosmology. A Planck maximum-likelihood cosmology yields spectra with elevated power relative to the fiducial D3A cosmology (and is thus in stronger tension with the measurements), whereas increasing the summed neutrino mass from 0.06 eV (fiducial) to 0.24 eV suppresses the power over all scales sampled here.  The LS8 `lensing' cosmology predicts the lowest power and is in best agreement with the tSZ power spectrum measurements.  \textit{Bottom left:} Dependence of the predicted tSZ power spectrum on baryon physics, namely variations in the gas fractions of groups and clusters which are mediated primarily through variations in the AGN feedback strength.    Feedback effects are most evident at small angular scales ($\ell \ga 2000$) and cannot reconcile the offset from the large-scale Planck measurements.  \textit{Bottom right:}  Dependence on other baryon variations, including variations in the stellar mass function (both at the fiducial and reduced gas fractions) and the calibrated and strong jet models of AGN feedback.   The effects of variations in the stellar mass function are generally negligible compared to those of variations in the gas fractions.
    }
    \label{fig:tSZ_powerspec}
\end{figure*}

In the top left panel of Fig.~\ref{fig:tSZ_powerspec} we examine the dependence of the tSZ effect power spectrum on simulation box size and resolution.   The solid coloured curves correspond to the predicted spectra for the \flamingo\ simulations with the fiducial D3A cosmology as the box size and resolution are varied.  We see that the simulation predictions are largely converged with resolution at fixed box size (1 Gpc), but the 1 Gpc boxes are missing a small amount of power compared to the 2.8 Gpc volume on all scales.  We attribute this offset, which is larger than for the case of cosmic shear, as being due to the disproportionate influence of very massive, rare clusters on this statistic.  Consistent with this interpretation is the relatively large degree of cosmic variance in the 2D map-based analysis (shaded blue region), which shows significant variation between the 8 lightcones, particularly for $\ell \la 300$.  We note, however, that the mean power spectrum from the 2D method (dashed blue curve) is generally in very good agreement with the fiducial 1D Limber calculation. 

Comparing the simulations to the observational measurements, it is immediately obvious from the top left panel of Fig.~\ref{fig:tSZ_powerspec} that the simulations predict far too much power on all scales relative to the Planck and SPT measurements.  However, as already noted, the tSZ effect power spectrum is very sensitive to the adopted cosmology.  In the top right panel of Fig.~\ref{fig:tSZ_powerspec} we therefore explore the cosmological variations in \flamingo.  A Planck maximum-likelihood cosmology yields spectra with slightly elevated power relative to the fiducial D3A cosmology (and is thus in stronger tension with the measurements), whereas increasing the summed neutrino mass from 0.06 eV (fiducial) to 0.24 eV suppresses the power over all scales sampled here, though not sufficiently to bring the simulations into agreement with the data.  The LS8 `lensing' cosmology predicts the lowest power and is in best agreement with the tSZ power spectrum measurements, although it still predicts slightly too much power relative to the Planck measurements at intermediate scales and particularly with respect to the SPT measurements on small angular scales.

In the bottom panels of Fig.~\ref{fig:tSZ_powerspec} we examine the feedback dependence of the tSZ effect.  The bottom left panel explores variations in the gas mass fractions of groups and clusters (again mediated primarily through AGN feedback variations), while the bottom right panel explores variations in the SMF (at the fiducial and reduced gas fractions) and variations in the nature of the AGN feedback implementation (thermal vs.~jet).  We conclude from these comparisons that the tSZ power spectrum is generally insensitive to even large variations in the baryon physics on the large angular scales probed by Planck, which is consistent with the findings of previous studies that used cosmological hydro simulations (e.g., \citealt{Battaglia2012,McCarthy2014}).  Note that while the feedback variations we have explored are generally unable to liberate baryons from the very massive haloes and large physical scales that dominate the tSZ effect, baryon physics can still in principle alter the power spectrum on large scales through variations in the efficiency of star formation, as star formation siphons off the hot gas reservoir that gives rise to the tSZ effect (see \citealt{DaSilva2001} for a dramatic example).  However, current observational measurements of the baryon fractions of massive clusters indicate that they have approximately their full cosmological complement of baryons (with $f_b \approx \Omega_b/\Omega_m$) and that the hot gas dominates (e.g., \citealt{Gonzalez2013,Akino2022}), implying that feedback has strongly curtailed star formation in these systems and their progenitors.  Such behaviour is also effectively enforced in the \flamingo\ simulations through calibration to the observed $z=0$ galaxy SMF.  We further highlight that the fiducial \flamingo\ simulation predicts a tSZ effect--halo mass scaling relation that is in excellent agreement with Planck tSZ cluster measurements (see figure 15 of \citealt{Schaye2023}).

The situation changes on the smaller scales probed by SPT (a few arcminutes), which are more sensitive to group-mass haloes.  Here the variations in feedback can give rise to relatively large effects on the tSZ power, although none of the variations we have explored here can reproduce the low amplitude of the SPT measurements for the fiducial cosmology.  It is possible that some combination of cosmological and feedback modifications (e.g., similar to the `lensing' cosmology but with stronger-than-fiducial feedback, or a lower amplitude cosmology with fiducial feedback) could reconcile these measurements, but we leave that as an open question for future work.

In summary, similar to the cosmic shear comparison in Section \ref{sec:shear-ps}, but with higher significance, the tSZ effect power spectrum prefers a low $S_8$ cosmology compared to the D3A and Planck cosmologies.  This conclusion is qualitatively consistent with some previous studies of the tSZ effect that also used cosmological hydrodynamical simulations (e.g., \citealt{McCarthy2014,McCarthy2018}). It is also consistent with the halo model-based analyses\footnote{We note that the tSZ power spectrum in \citet{Planck2016_tSZ_PS} is of somewhat higher amplitude than in the re-analysis by \citet{Bolliet2018}, which is likely the reason why the inferred value of $S_8$ from the former study is slightly larger than one would anticipate based on the comparison in Fig.~\ref{fig:tSZ_powerspec}.  In addition, the uncertainties are larger in the latter study due to the inclusion of the tri-spectrum in the covariance matrix.  See \citet{Bolliet2018} for further discussion.} of \citet{Planck2014,Planck2016_tSZ_PS} and \citet{Bolliet2018} who infer $S_8 \approx 0.78 \pm 0.02$ and $S_8 \approx 0.75 \pm 0.04$, respectively, when adopting a hydrostatic mass bias consistent with weak lensing observations and the predictions of simulations. (By contrast, \citealt{Reichardt2021} require $S_8 \approx 0.69 \pm 0.03$ to match their SPT measurements when using the halo model of \citealt{Shaw2010}, although as already noted these scales can be significantly affected by feedback.)  Note, however, that in the context of the halo model one can boost the best-fit $S_8$ by appealing to a larger halo mass bias, but only at the expense of agreement with simulation predictions and observational weak lensing mass constraints.  One advantage of the comparison in the present study is that we go directly from either 3D power spectra or 2D maps to a tSZ power spectrum prediction without the intermediate step of defining and counting haloes and choosing a mass bias.  The comparison here is therefore more direct.  

While the discussion of the impact of baryon feedback on current cosmic shear power spectrum constraints remains an open discussion due to the degeneracy between cosmology and baryon feedback (e.g., \citealt{Schneider2022,Troster2022,Amon2022,Chen2023,Arico2023}), it is much more difficult to appeal to baryons as a solution to the tSZ effect power offset on large scales, owing to the fact that this statistic is dominated by very massive clusters which are observed to be `baryonically closed' and dominated by hot gas.

\begin{figure}
    \includegraphics[width=\columnwidth]{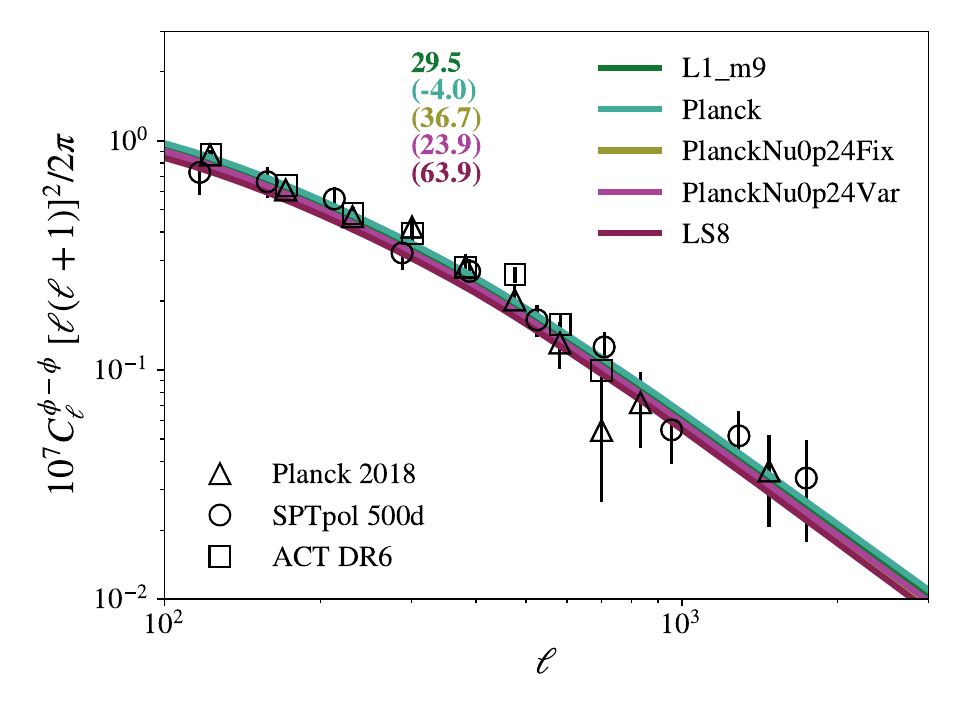}
    \caption{Dependence of the CMB lensing angular power spectrum on cosmology.  The open triangles, circles, and squares, correspond to the Planck 2018, SPTpol 500d, and ACT DR6 measurements of \citet{Planck_lensing}, \citet{Wu2019} and \citet{Qu2023}, respectively.  The error bars correspond to the diagonal components of the covariance matrices.  Note that the $\chi^2$ values displayed in the top left are computed by summing over the three data sets and take into account only the diagonal errors.
    A Planck maximum-likelihood cosmology yields spectra with slightly elevated power relative to the fiducial D3A (with a similarly good match to the measurements), whereas increasing the summed neutrino mass from 0.06 eV (fiducial) to 0.24 eV suppresses the power and worsens the agreement with the data.  The LS8 `lensing' cosmology predicts the lowest power, but in contrast with galaxy lensing and the tSZ effect tests, this model yields the worst agreement with the CMB lensing measurements.}
    \label{fig:CMB_powerspec}
\end{figure}

\subsection{CMB lensing power spectrum}
\label{sec:CMB-ps}

The final auto power spectrum that we consider is the CMB lensing power spectrum.  As already noted in Section \ref{sec:obs}, CMB lensing receives contributions from matter fluctuations over a wide range of redshifts, with a $C_\ell$-weighted mean redshift of $z_{\rm eff} \approx 3-4$ (see Fig.~\ref{fig:keff_zeff}).  For a review of CMB lensing, see \citet{Lewis2006}.  Current measurements of this statistic, which are essentially derived from four-point measurements of the CMB temperature and polarisation maps \citep{Hu2002}, are generally restricted to relatively large angular scales of $\ell \la 2000$.   Given that much of the signal arises from high redshifts, existing measurements typically probe linear scales, which considerably simplifies the modelling the CMB lensing power spectrum.  Indeed, in Appendix \ref{sec:other} we confirm that this statistic is insensitive to variations in box size, resolution, and baryon physics.  This situation will change in the near future, however, as forthcoming CMB lensing experiments such as the Simons Observatory \citep{Ade2019} and CMB-S4 \citep{Abazajian2019} will begin to probe scales that are more sensitive to non-linear evolution and baryonic physics \citep{McCarthy2022,Upadhye2023}.  Here we focus on the cosmological dependence of the CMB lensing power spectrum.

In Fig.~\ref{fig:CMB_powerspec} we explore the cosmology variations in \flamingo\ and compare with the latest measurements from Planck 2018 \citep{Planck_lensing}, SPTpol 500d \citep{Wu2019}, and ACT DR6 \citep{Qu2023}.  The quoted $\chi^2$ values sum over the three data sets and are calculated with respect to the diagonal elements only of the respective covariance matrices.  While the full covariance matrices are available, combining them in a rigorous way is non-trivial due to the spatial overlap of the surveys.  Furthermore, as the CMB lensing power spectrum probes mainly linear scales, the off-diagonal uncertainties are expected to be small.  Indeed, \citet{Qu2023} find the off-diagonal correlations to typically be less than 10\% of the total uncertainty.

In agreement with the above studies, we find that a Planck-like cosmology yields an excellent match to the CMB lensing measurements.  The Planck maximum-likelihood cosmology yields spectra with slightly elevated power relative to the fiducial D3A, but with a similarly good match to the measurements.  Increasing the summed neutrino mass from 0.06 eV (fiducial) to 0.24 eV suppresses the power and worsens the agreement with the observed power spectrum.  This is an important result, since one of the proposed ways of reconciling the primary CMB fluctuations with the apparent suppression of the growth of LSS is to appeal to the fact that neutrinos do not cluster significantly on small scales.  Some previous studies have suggested values of $\sum m_\nu \approx 0.2-0.4$ eV (e.g., \citealt{Battye2014,Beutler2014,Wyman2014,McCarthy2018}) could reconcile most of the tension.  However, here we see that raising the summed mass of neutrinos worsens the agreement with the observed CMB lensing power spectrum, by suppressing the predicted amplitude below what is measured.  Thus, CMB lensing observations play a critical role in challenging massive neutrinos as a solution to the $S_8$ tension.  Lastly, the LS8 `lensing' cosmology predicts the lowest power and in contrast with cosmic shear and the tSZ effect tests discussed above, this model yields the worst agreement with the CMB lensing measurements.

In the most recent analysis using data from ACT DR6, \citet{Madhavacheril2023} find that the combination of ACT lensing + Planck lensing + BAO yields a tight constraint of $S_8 = 0.83 \pm 0.02$, in excellent agreement with the Planck CMB cosmology and the fiducial D3A cosmology.  Note that CMB lensing alone actually best constrains the parameter combination $\sigma_8 \Omega_m^{0.25}$ (rather than $S_8$) and a fit to the ACT DR6 measurements yields $0.61 \pm 0.02$ for this quantity \citep{Qu2023}.  For context, our fiducial D3A, Planck, and LS8 cosmologies have $\sigma_8 \Omega_m^{0.25} = 0.60$, $0.61$, and $0.56$.  Thus, CMB lensing alone agrees extremely well with the Planck CMB and D3A cosmologies and is in a moderate degree of tension with the LS8 cosmology.  Note that the two 0.24 eV neutrino cosmologies yield values of $0.61$ and $0.59$ for the Var and Fix variants, respectively, which is consistent with the CMB lensing-only constraints of \citep{Qu2023}.  However, it is clear from the $\chi^2$ values in Fig.~\ref{fig:CMB_powerspec} that the neutrino cosmologies fit the measurements worse than the D3A and Planck cosmologies, which may imply that the $\sigma_8 \Omega_m^{0.25}$ parameter does not capture the full cosmological dependence of the CMB lensing-only constraints.

Assuming no significant systematic errors have been neglected in the three auto-power spectra comparisons above (either on the observational or theoretical sides) which would alter the conclusions drawn, then based on Fig.~\ref{fig:keff_zeff} there are two possible generic ways to reconcile the three probes explored so far: i) a modification to the redshift evolution of matter fluctuations, such that the fluctuations grow more slowly at late times than predicted in the fiducial cosmology (noting that CMB lensing probes higher redshifts than cosmic shear or the tSZ effect; see Fig.~\ref{fig:keff_zeff}, bottom panel); and/or ii) a modification on non-linear scales, such that the fluctuations on these scales grow more slowly than expected (cosmic shear and tSZ effect probe non-linear scales, whereas CMB lensing probes linear scales; see Fig.~\ref{fig:keff_zeff}, top panel).  \citet{Preston2023} (see also \citealt{Nguyen2023}) recently came to the same general conclusions.  Here, however, we argue using the \flamingo\ hydrodynamical simulations that baryonic feedback is unlikely to be the physics driving the required modifications.

\subsection{Cosmic shear -- tSZ effect cross-spectrum}
\label{sec:shear-tSZ-ps}

Having examined the auto power spectra of the three observables (shear, tSZ, CMB lensing) in the previous sections, we now examine the cross-spectra between these variables.  Importantly, the cross-spectra contain additional information about these observables.  For example, current measurements of the cosmic shear power spectrum are most sensitive to the clustering (2-halo) and structure (1-halo) of group-mass haloes ($M \sim 10^{13-14}$ M$_\odot$) at a distance roughly half way to the background source population, whereas the tSZ effect power spectrum is mostly sensitive to the structure of low-redshift, very massive haloes ($M \sim 10^{15}$ M$_\odot$; \citealt{Komatsu2002}).  The cross-spectrum, however, is sensitive to both the clustering and structure of haloes with redshifts and masses intermediate between these regimes (i.e., $M \sim 10^{14-15}$ M$_\odot$), as shown in \citet{Mead2020}.  This is simply because the cross-spectrum picks out (only those) structures that contribute significantly to both observables.

\begin{figure*}
    \includegraphics[clip, trim=3.5cm 3.5cm 3.5cm 4.0cm, width=\textwidth]{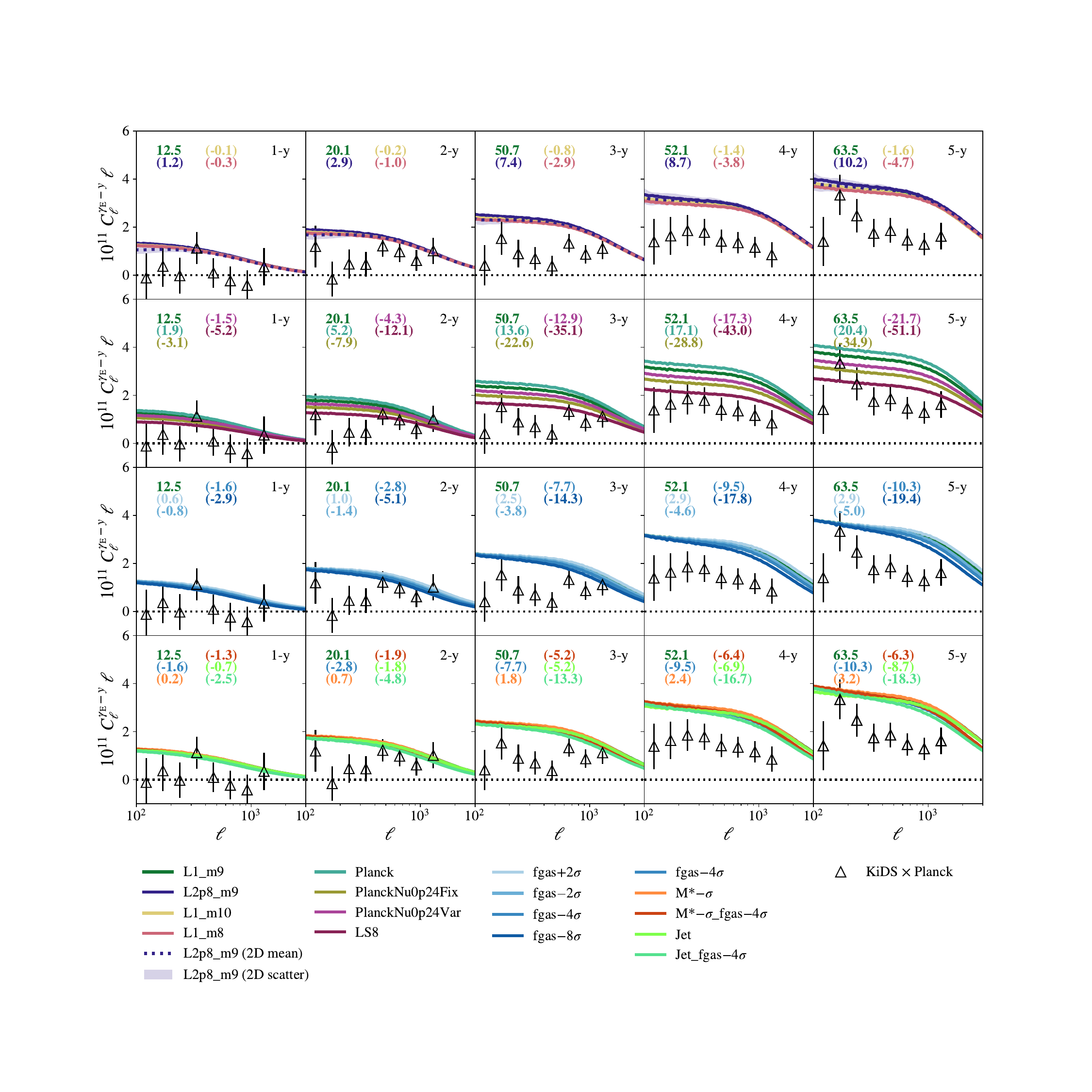}
    \caption{The cosmic shear--tSZ effect angular cross-power spectrum.  The open triangles correspond to the KiDS1000 $\times$ Planck measurements of \citet{Troster2022}.  The different columns correspond to the cross-spectrum between different KiDS tomographic bins (1-5) and the tSZ data.  \textit{Top row:} Dependence on simulation box size and resolution.  The simulation predictions are largely converged with resolution, though the 1 Gpc boxes are missing a small amount of power compared to the 2.8 Gpc volume. \textit{Second row:} Dependence on cosmology.  A Planck maximum-likelihood cosmology yields spectra with elevated power relative to the fiducial D3A (and is thus in stronger tension with the measurements), whereas increasing the summed neutrino mass from 0.06 eV (fiducial) to 0.24 eV suppresses the power over all scales sampled here.  The LS8 `lensing' cosmology predicts the lowest power and is in best agreement with the measurements, though it still predicts slightly too much power.  \textit{Third row:} Dependence on baryon physics, namely variations in the gas fractions of groups and clusters which are mediated primarily through variations in the AGN feedback strength.   Feedback effects are significant on scales of $\ell \ga 500$ but cannot reconcile the offset at larger scales.  \textit{Bottom row:}  Dependence on other baryon variations, including variations in the stellar mass function (both at the fiducial and reduced gas fractions) and the fiducial and strong jet models of AGN feedback.  The effects of variations in the stellar mass function are generally negligible compared to those of variations in the gas fractions.}
    \label{fig:tSZ_shear_powerspec}
\end{figure*}

Another important benefit of cross-spectra measurements is that they have different biases than auto power spectra.  For example, the noise (e.g., shot noise, detector noise) in two independent maps, such as cosmic shear and tSZ, is expected to be uncorrelated and therefore will not contribute to the cross-spectrum between those maps, whereas if unaccounted for, noise can significantly bias both observational and simulation measurements of auto power spectra.  In the case of simulations, for example, particle shot noise must be subtracted from the power spectra.  Other forms of bias (e.g., in galaxy shape estimation, the separation of the tSZ from the clustered infrared background, etc.) will enter into the auto power spectrum differently than for the cross-spectrum.  For example, multiplicative forms of bias will enter into the power spectrum squared, but only linearly in the cross-spectrum.  Thus, a simultaneous examination of the auto and cross-power spectra gives an important cross-check on possible biases that could be affecting both measurements.

In the context of the impact of baryons on large-scale structure, the cosmic shear--tSZ effect cross-spectrum is also interesting for another reason.  Specifically, this cross-spectrum is essentially a measurement of how the hot gas (in particular its thermal energy density) traces that of the underlying matter field.  Thus, in addition to yielding another cosmological test, this statistic also provides a valuable opportunity to assess the realism of feedback models when it is examined on small scales, as feedback is known to strongly alter the hot gas properties of groups and clusters (e.g., \citealt{McCarthy2010,Planelles2014,LeBrun2014,Henden2018,Oppenheimer2021}). 

\citet{Troster2022} recently performed a spatial cross-correlation analysis of the KiDS 1000 tomographic data set with tSZ effect maps constructed from the Planck 2015 data set \citep{Planck2016_tSZ_PS} and the ACT DR4 data set \citep{MallabyKay2021}.  The ACT data have lower noise than the Planck $y$ map, but the overlap between the ACT DR4 and KiDS 1000 surveys is only partial.  Thus, in practice the cross-spectrum derived using the Planck data set is better constrained and we therefore focus our comparison on that data set.  Note that in the comparisons presented in the previous sections, the effects of the beam and pixel window function were deconvolved from the observed power spectra, whereas \citet{Troster2022} have not deconvolved either effect (electing instead to convolve the theory).  For consistency with the previous comparisons, we deconvolve the Planck beam and the $N_{\rm side}=2048$ pixel window functions from the measured cross-spectra of \citet{Troster2022}. 

In Fig.~\ref{fig:tSZ_shear_powerspec} we present a comparison of the full suite of \flamingo\ simulations with the measurements of \citet{Troster2022}, processed as described above.  The columns are organised by cosmic shear tomographic bin (1 - 5; low redshift to high redshift from left to right).  The rows follow our previous comparisons (from top to bottom): i) impact of simulation resolution, box size, and analysis method (1D vs. 2D); ii) dependence on cosmology; iii) dependence on group/cluster gas fraction; and iv) other baryonic variations.

In the top row of Fig.~\ref{fig:tSZ_shear_powerspec} we see that the predicted cross-spectra are relatively well converged with resolution at fixed box size (1 Gpc).  The high-resolution simulation (L1\_m8) predicts slightly less power than the fiducial (L1\_m9) and low res (L1\_m10) runs, owing to a slightly higher star formation efficiency in groups/clusters in the high res. simulation (this was also visible in the tSZ effect power spectrum in Fig.~\ref{fig:tSZ_powerspec}).  Comparing the 1 Gpc and 2.8 Gpc fiducial resolution runs, the former are missing a small amount of power on the largest angular scales, which is most evident for the cross spectra involving the higher tomographic bins.  Similar to the tSZ effect power spectrum, we attribute this slight offset as being due to a non-negligible contribution of massive, rare haloes.  There is good agreement between the fiducial 1D Limber methodology (solid curves) and the map-based 2D analysis (dashed curve), the latter of which shows that cosmic variance becomes relevant on scales of $\ell \la 500$ (shaded region).   It is also clear from the top row of Fig.~\ref{fig:tSZ_shear_powerspec} that the fiducial calibrated \flamingo\ simulations in the fiducial D3A cosmology predicts too much power compared to the observational measurements on all but the smallest scales.  This is particularly evident for the highest tomographic bins.  

\begin{table}
	\centering
	\caption{$\chi^2$ values of the predicted KiDS 1000 cosmic shear--Planck tSZ cross-power spectra with respect to the measurements of \citet{Troster2022}, for which there are 40 independent data points summed over the 5 cross-spectra.  Values in parentheses indicate the $\Delta \chi^2$ with respect to the L1\_m9 run.  $\chi^2$ values using either the diagonal elements alone or the full covariance matrix are provided.}
	\label{tab:shear_y_chi2}
	\begin{tabular}{lrr} 
		\hline
		Prefix & $\chi^2_{\rm diag}$ & $\chi^2_{\rm covar}$ \\
		\hline
L1$\_$m9 & 198.9 & 78.9\\
L2p8$\_$m9 & (30.4) & (7.7)\\
L1$\_$m10 & (-4.1) & (-2.0)\\
L1$\_$m8 & (-12.7) & (-3.0)\\
fgas$+2\sigma$ & (9.9) & (3.9)\\
fgas$-2\sigma$ & (-15.5) & (-5.5)\\
fgas$-4\sigma$ & (-31.8) & (-11.0)\\
fgas$-8\sigma$ & (-59.5) & (-19.9)\\
M*$-\sigma$ & (8.3) & (1.7)\\
M*$-\sigma$\_fgas$-4\sigma$ & (-21.0) & (-8.3)\\
Jet & (-23.4) & (-6.5)\\
Jet\_fgas$-4\sigma$ & (-55.6) & (-18.8)\\
Planck & (58.1) & (16.2)\\
PlanckNu0p24Fix  & (-97.4) & (-26.9)\\
PlanckNu0p24Var & (-57.7) & (-16.3)\\
LS8 & (-146.5) & (-39.8)\\
            \hline
	\end{tabular}
\end{table}

In the second row from the top, we examine the cosmology dependence of the shear--tSZ cross.  Here we see that increasing the neutrino mass, whose main effect is to lower $S_8$ (or, more generally, the clustering amplitude), or directly lowering $S_8$ at a fixed minimal neutrino mass (LS8), lowers the amplitude of the predicted cross-spectra, yielding an improved match to the data.  Consistent with the cosmic shear and tSZ effect power spectrum comparisons in Sections \ref{sec:shear-ps} and \ref{sec:tSZ-ps}, respectively, we find the `lensing' cosmology (LS8) yields the best agreement with the data, although it is still somewhat elevated with respect to the observations (similar to the tSZ effect power spectrum in Fig.~\ref{fig:tSZ_powerspec}).  The Planck cosmology, on the other hand, yields a higher amplitude than the fiducial D3A cosmology and is therefore in slightly stronger tension with the observational measurements.  Note that the shear--tSZ cross spectrum best constrains a combination of $\sigma_8$ and $\Omega_m$ that differs from the definition of $S_8$.  \citet{Troster2022} find that the parameter $\Sigma_8^{0.2} \equiv \sigma_8 (\Omega_m/0.3)^{0.2}$ describes the degeneracy well for the shear--tSZ cross.  They determine $\Sigma_8^{0.2} \approx 0.72 \pm 0.04$.  For comparison, the D3A, Planck, and LS8 cosmologies used here have $\Sigma_8^{0.2} = 0.81$, $0.82$, and $0.76$, respectively.  Because cosmic shear and the shear--tSZ cross have different dependencies on $\Omega_m$ and $\sigma_8$, a joint analysis helps to break the degeneracy between these parameters (see figure 6 of \citealt{Troster2022}; see also \citealt{Fang2023}).  As already mentioned in Section \ref{sec:shear-ps}, jointly modelling these two probes, \citet{Troster2022} find $S_8 \approx 0.75 \pm 0.02$, representing a $\approx 3\sigma$ tension with the Planck CMB cosmology.

In the bottom two rows of Fig.~\ref{fig:tSZ_shear_powerspec} we examine the feedback dependence of the shear--tSZ cross spectrum.  Variations in feedback lead to noticeable differences in the predictions on scales of $\ell \ga 400$ and the spread in the predictions becomes comparable to the spread due to cosmological variations on scales of $\ell \ga 2000$, making this statistic more sensitive to feedback variations than the cosmic shear power spectrum on these scales.  Nevertheless, on the basis of Fig.~\ref{fig:tSZ_shear_powerspec} we generally conclude that the feedback variations we have explored cannot reconcile the offset between the data and the fiducial D3A cosmology, as many of the bins are on angular scales that are not significantly impacted by baryons.  This is consistent with the findings of \citet{Troster2022} who, even though the impact of baryons has been marginalised over using the halo model of \citet{Mead2020}, still find a $\approx 3\sigma$ tension with the Planck CMB cosmology, similar to that derived from cosmic shear alone.  

\begin{figure*}
    \includegraphics[width=\columnwidth]{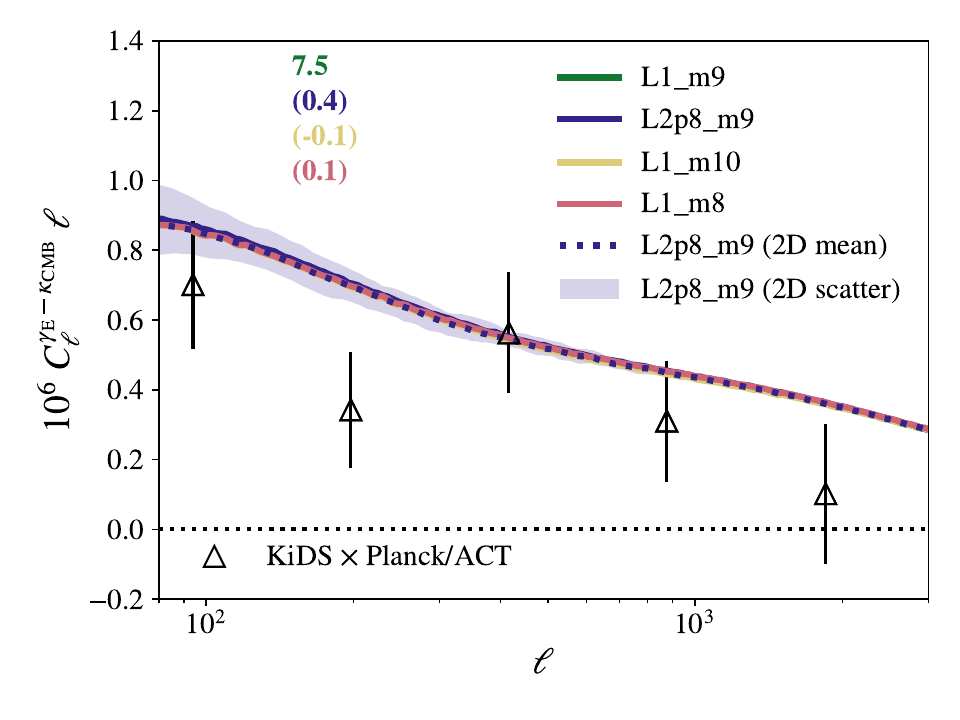}
    \includegraphics[width=\columnwidth]{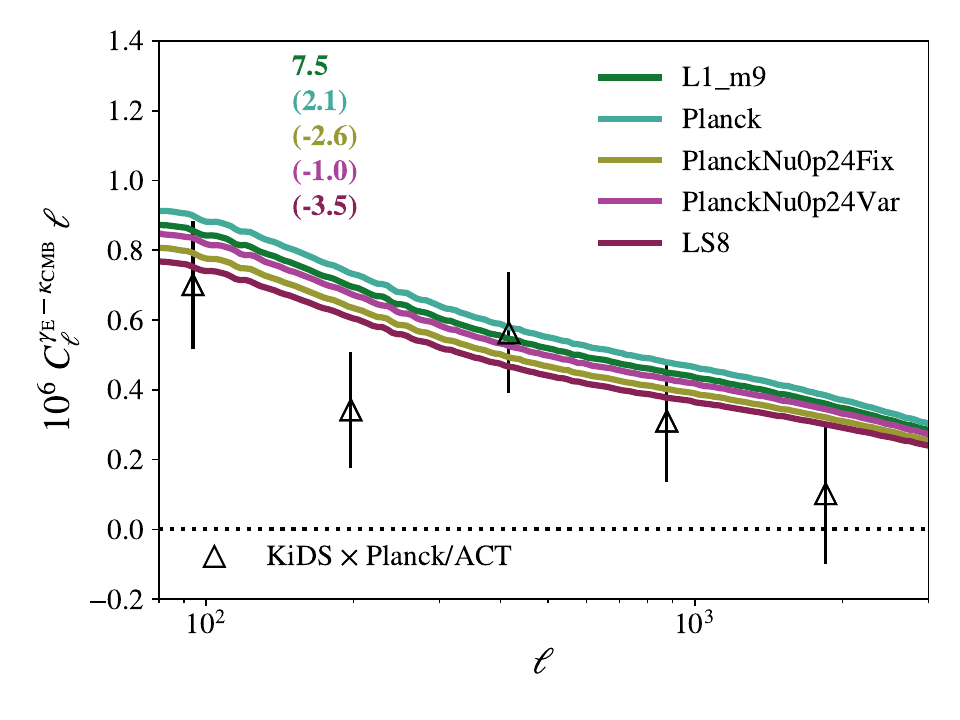}\\
    \includegraphics[width=\columnwidth]{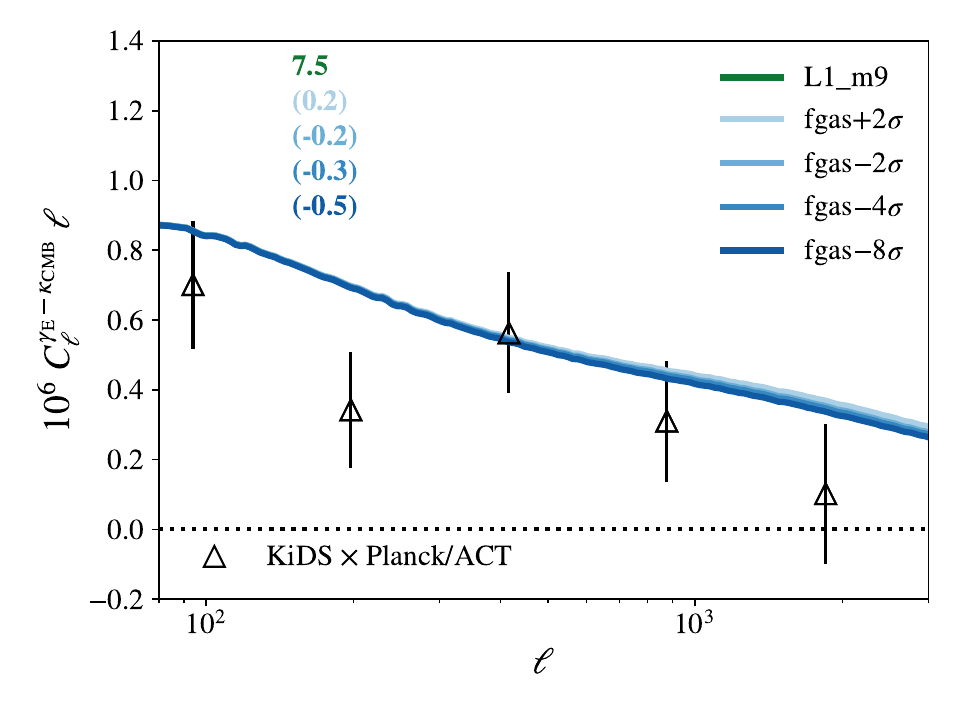}
    \includegraphics[width=\columnwidth]{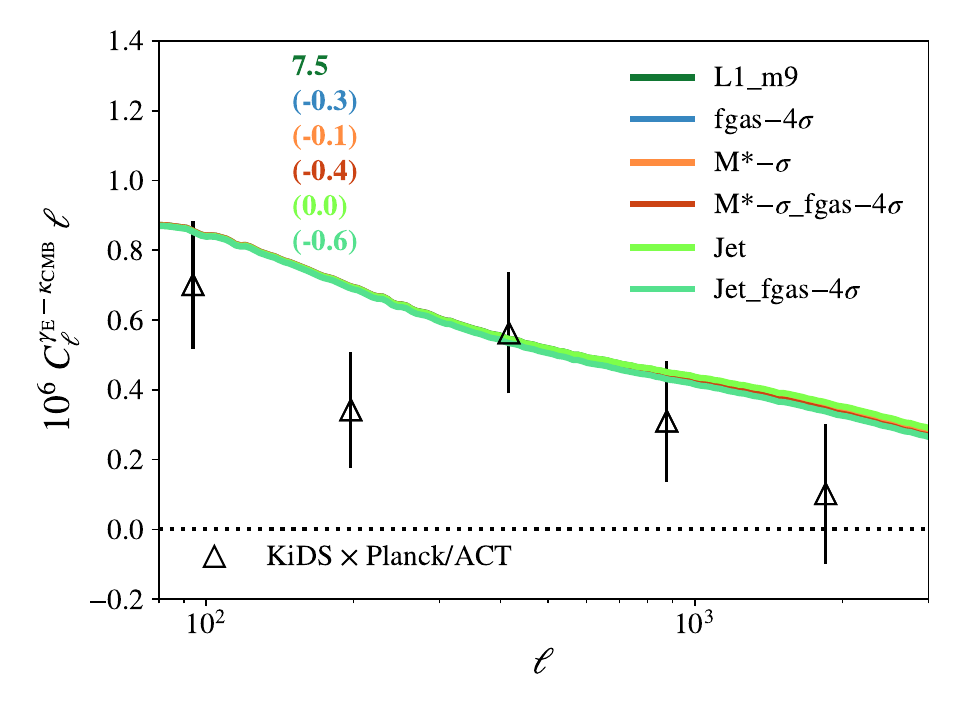}
    \caption{The cosmic shear--CMB lensing angular cross-spectrum.  The open triangles correspond to the KiDS 1000 $\times$ Planck/ACT CMB lensing cross-spectrum measurements of  \citet{Robertson2021}.  The $\chi^2$ values displayed in the top left of each panel take into account only the diagonal elements of the covariance matrix. \textit{Top left:} Dependence on simulation resolution and box size.  The simulation predictions are converged with resolution, box size, and analysis method (1D; solid curves vs. 2D; dashed curve).  The shaded region shows the scatter between the different light cones for the 2D map-based method.  \textit{Top right:} Dependence on cosmology.   A Planck maximum-likelihood cosmology yields spectra with elevated power relative to the fiducial D3A (and is thus in stronger tension with the measurements), whereas increasing the summed neutrino mass from 0.06 eV (fiducial) to 0.24 eV suppresses the power over all scales sampled here.  The LS8 `lensing' cosmology predicts the lowest power and is in best agreement with the cross power spectrum measurements.  \textit{Bottom left:} Dependence on baryon physics, namely variations in the gas fractions of groups and clusters which are mediated primarily through variations in the AGN feedback strength.  Feedback effects are most evident at small angular scales ($\ell \ga 1000$) and cannot reconcile the offset from the large-scale measurements.  \textit{Bottom right:}  Dependence on other baryon variations, including variations in the stellar mass function (both at the fiducial and reduced gas fractions) and the fiducial and strong jet models of AGN feedback.  The effects of variations in the stellar mass function are generally negligible compared to variations in the gas fractions.}
    \label{fig:shear_CMB_powerspec}
\end{figure*}

In Table \ref{tab:shear_y_chi2} we list the $\chi^2$ values of the cosmic shear--tSZ cross power spectra.  The values are summed over the 5 cross-spectra and are listed for both the cases where only the diagonal errors are adopted and when the full covariance matrix is used.  The tabulated values confirm the visual trends in Fig.~\ref{fig:tSZ_shear_powerspec}, although the differentiability of the models is reduced when taking into account the full covariance of the measurements.

In Appendix \ref{sec:other} we present a comparison to an independent measurement of the shear--tSZ cross spectrum from \citet{Hojjati2017}.  Those authors performed a spatial cross-correlation analysis between the Planck 2015 tSZ map and cosmic shear data from the RCSLenS survey \citep{Hildebrandt2016}.  While the KiDS 1000 survey covers a larger area and is significantly deeper than RCSLenS, it is nevertheless useful to check whether the conclusions are consistent.  In short, as shown in Fig.~\ref{fig:tSZ_shear_RCS_powerspec}, four of the five multipole bins from \citet{Hojjati2017} have amplitudes lower than predicted by the fiducial \flamingo\ model in the fiducial D3A cosmology, in general agreement with the findings above.  However, these also correspond to relatively small angular scales of $\ell \sim 1000$ where the impact of feedback is non-negligible.  Thus, while there is general consistency with the KiDS 1000 $\times$ Planck tSZ results above, the significance of the tension with the D3A cosmology is less compelling for the RCSLenS-based comparison owing to the uncertainties of baryonic modelling. 

\subsection{Cosmic shear -- CMB lensing cross-spectrum}
\label{sec:shear-CMB-ps}

The final statistic we examine is the cross-spectrum between cosmic shear and CMB lensing.  Specifically, in Fig.~\ref{fig:shear_CMB_powerspec} we compare the predictions of the \flamingo\ simulations to the recent measurements of \citet{Robertson2021}.  Those authors performed a spatial cross-correlation analysis between the KiDS 1000 cosmic shear data and the CMB lensing maps from the Planck 2018 and ACT DR4 releases.  We use their combined Planck/ACT measurements and uncertainties.  The uncertainties correspond to the diagonal elements of the covariance matrix only, as the full covariance matrix is not publicly available.  Note that \citet{Robertson2021} used a single large tomographic bin for KiDS data ($0.1 < z_{\rm B} < 1.2$) and we use the source redshift distribution, $n(z)$, shown in figure 1 of their paper.  As we demonstrated in Fig.~\ref{fig:keff_zeff} of the present study, this particular cross-spectrum generally probes larger scales and higher redshifts than cosmic shear alone.  We find that the measurements of \citet{Robertson2021} are sensitive to both linear scales ($k \approx 0.1$ [$h$/Mpc]) at $\ell \approx 100$ and non-linear scales ($k \approx 3$ [$h$/Mpc]) at $\ell \approx 2000$, with a mean ($C_\ell$-weighted) redshift of $z_{\rm eff} \approx 0.5$.

\begin{table*}
	\centering
	\caption{The $\chi^2$ values and number of standard deviations, $N_{\sigma}$, of each simulation with respect to the observed auto- and cross-power spectra.  For the cosmic shear power spectrum ($\gamma_\textrm{E}$-$\gamma_\textrm{E}$) and cosmic shear--tSZ effect cross-spectrum ($\gamma_\textrm{E}$-$y$) we use the full covariance matrices to compute $\chi^2$, whereas for the other cases we use the diagonal errors only.  We compute $\chi^2$ values with respect to the following observational data sets: the KiDS 1000 cosmic shear measurements of \citet{Troster2022} [$\gamma_\textrm{E}$-$\gamma_\textrm{E}$ (KiDS)], the DES Y3 cosmic shear measurements of \citet{Doux2022} [$\gamma_\textrm{E}$-$\gamma_\textrm{E}$ (DES)], the Planck tSZ effect power spectrum measurements of \citet{Bolliet2018} [$y$-$y$], the Planck+SPT tSZ effect power spectrum measurements of \citet{Bolliet2018} and \citet{Reichardt2021} [$y$-$y$ (+SPT)], the Planck, SPT, and ACT CMB lensing power spectrum measurements of \citet{Planck_lensing}, \citet{Wu2019}, and \citet{Qu2023}, respectively [$\kappa_\textrm{CMB}$-$\kappa_\textrm{CMB}$], the KiDS 1000 cosmic shear--Planck tSZ effect cross-spectrum of \citet{Troster2022} [$\gamma_\textrm{E}$-$y$], the KiDS 1000 cosmic shear--ACT/Planck CMB lensing cross-spectrum of \citet{Robertson2021} [$\gamma_\textrm{E}$-$\kappa_\textrm{CMB}$], and the Planck CMB lensing--tSZ effect cross-spectrum of \citet{Hill2014} [$\kappa_\textrm{CMB}$-$y$].  The numbers displayed in bold face correspond to the models that are within $3\sigma$ of the observational measurements.}
	\label{tab:chi2_nsig}
        \begin{tabular}{lrrrrrrrrrrrrrrrr}
                   & \multicolumn{2}{c}{$\gamma_\textrm{E}$-$\gamma_\textrm{E}$ (KiDS)} & 
                   \multicolumn{2}{c}{$\gamma_\textrm{E}$-$\gamma_\textrm{E}$ (DES)} &
                   \multicolumn{2}{c}{$y$-$y$} & \multicolumn{2}{c}{$y$-$y$ (+SPT)} & \multicolumn{2}{c}{$\kappa_\textrm{CMB}$-$\kappa_\textrm{CMB}$} 
                   & \multicolumn{2}{c}{$\gamma_\textrm{E}$-$y$} & \multicolumn{2}{c}{$\gamma_\textrm{E}$-$\kappa_\textrm{CMB}$} & \multicolumn{2}{c}{$\kappa_\textrm{CMB}$-$y$}\\
                \hline
		Prefix & $\chi^2$ & $N_{\sigma}$ & $\chi^2$ & $N_{\sigma}$ & $\chi^2$ & $N_{\sigma}$ & $\chi^2$ & $N_{\sigma}$ & $\chi^2$ & $N_{\sigma}$ & $\chi^2$ & $N_{\sigma}$ & $\chi^2$ & $N_{\sigma}$ & $\chi^2$ & $N_{\sigma}$\\
		\hline
              L1$\_$m9 & 183.2 & 3.6 & 324.3 & \textbf{2.3} & 139.7 & 10.5 & 435.3 & 19.9 & 29.5 & \textbf{0.4} & 78.9 & 3.6 & 7.5 & \textbf{1.2} & 2.8 & \textbf{0.2}\\
            L2p8$\_$m9 & 185.4 & 3.7 & 327.7 & \textbf{2.4} & 225.7 & 13.9 & 568.4 & 22.9 & 27.1 & \textbf{0.1} & 86.6 & 4.1 & 8.0 & \textbf{1.3} & 3.2 & \textbf{0.4}\\
            L1$\_$m10 & 181.5 & 3.5 & 321.5 & \textbf{2.2} & 160.0 & 11.4 & 455.3 & 20.4 & 29.8 & \textbf{0.5} & 76.9 & 3.5 & 7.4 & \textbf{1.2} & 2.8 & \textbf{0.2}\\
            L1$\_$m8 & 184.6 & 3.7 & 325.5 & \textbf{2.3} & 115.2 & 9.4 & 372.0 & 18.2 & 29.4 & \textbf{0.4} & 75.9 & 3.4 & 7.7 & \textbf{1.3} & 2.6 & \textbf{0.1}\\
            fgas$+2\sigma$ & 185.6 &  3.7 & 325.5 & \textbf{2.3} & 140.4 & 10.6 & 470.6 & 20.7 & 29.4 & \textbf{0.4} & 82.8 & 3.9 & 7.7 & \textbf{1.3} & 2.7 & \textbf{0.2}\\
            fgas$-2\sigma$ & 181.2  & 3.5 & 323.3 & \textbf{2.2} & 129.0 & 10.0 & 368.0 & 18.1 &  29.5 & \textbf{0.4} & 73.5 & 3.2 & 7.4 & \textbf{1.2}& 2.8 & \textbf{0.2}\\
            fgas$-4\sigma$ & 179.4  & 3.4 & 322.5 & \textbf{2.2} & 117.2 & 9.5 & 302.8 & 16.3 & 29.6 & \textbf{0.4} & 67.9 & \textbf{2.8} & 7.2 & \textbf{1.1} & 2.8 & \textbf{0.2}\\
            fgas$-8\sigma$ & 176.6  & 3.3 & 321.6 & \textbf{2.2} & 91.5 & 8.1 & 198.4 & 12.8 & 29.7 & \textbf{0.4} & 59.0 & \textbf{2.0} & 7.0 & \textbf{1.1} & 3.0 & \textbf{0.3}\\
            M*$-\sigma$ & 181.9 & 3.6 & 323.8 & \textbf{2.2} & 161.0 & 11.5 & 468.9 & 20.7 & 29.5 & \textbf{0.4} & 80.7 & 3.7 & 7.4 & \textbf{1.2} & 3.1 & \textbf{0.3}\\
            M*$-\sigma$\_fgas$-4\sigma$ & 178.6 & 3.4 & 322.6 & \textbf{2.2} & 137.2 & 10.4 & 333.1 & 17.2 & 29.6 & \textbf{0.4} & 70.6 & 3.0 & 7.2 & \textbf{1.1} & 3.2 & \textbf{0.4} \\
            Jet & 183.8 & 3.7 & 323.6 & \textbf{2.2} & 106.7 & 8.9 & 327.4 & 17.1 & 29.6 & \textbf{0.4} & 72.4 & 3.1 & 7.6 & \textbf{1.2} & 2.3 & \textbf{0.0}\\
            Jet\_fgas$-4\sigma$ & 176.1 & 3.3 & 320.1 & \textbf{2.1} & 97.8 & 8.5 & 231.1 & 14.1 & 29.8 & \textbf{0.5} & 60.1 & \textbf{2.1} & 6.9 & \textbf{1.1} & 3.4 & \textbf{0.4}\\
            Planck & 199.3 & 4.4 & 348.5 & 3.2 & 187.0  & 12.5 & 552.9 & 22.6 & 25.4 & \textbf{0.1} & 95.2 & 4.7 & 9.6 & \textbf{1.7} & 3.8 & \textbf{0.6}\\
            PlanckNu0p24Fix & 166.9 & \textbf{2.8} & 311.2 & \textbf{1.8} & 47.4 & 5.2 & 184.4 & 12.3 & 66.2 & 4.0 & 52.0 & \textbf{1.4} & 5.0 & \textbf{0.6} & 1.8 & \textbf{0.3}\\
            PlanckNu0p24Var & 176.9 & 3.3 & 316.8 & \textbf{2.0} & 75.0 & 7.1 & 258.1 & 14.9 & 53.4 & \textbf{2.9} & 62.6 & \textbf{2.3} & 6.5 & \textbf{1.0} & 1.9 & \textbf{0.2}\\
            LS8 & 162.0 & \textbf{2.6} & 315.7 & \textbf{1.9} & 8.8 & \textbf{0.4} & 63.2 & 6.2 & 93.4 & 5.8 & 39.2 & \textbf{0.1} & 4.0 & \textbf{0.2} & 2.4 & \textbf{0.0}\\
		\hline
 \end{tabular}
\end{table*}

In the top left panel of Fig.~\ref{fig:shear_CMB_powerspec} we examine the dependence of the predicted shear--CMB lensing cross-spectrum on simulation box size and resolution, as well as analysis method (1D vs.~2D).  We find this cross-spectrum is converged with respect to box size and resolution.  Furthermore, there is excellent consistency between the fiducial 1D Limber method and the 2D map-based method for analysing the simulations.  The map-based analysis demonstrates that there is considerable cosmic variance between the different light cones on angular scales of $\ell \la 500$, owing to the fact that this corresponds to large physical scales which are sampled with relatively few modes even in the large 2.8 Gpc box.  As highlighted previously, the simulation ICs employ mode fixing on large physics scales and therefore the fiducial 1D Limber predictions ought to predict the mean shear--CMB lensing power spectrum to high accuracy for our chosen cosmology.

As in the previous comparisons (with the clear exception of the CMB lensing power spectrum in Section \ref{sec:CMB-ps}), the predicted shear--CMB lensing cross-spectrum for the fiducial calibrated \flamingo\ hydro simulation in the fiducial D3A cosmology has too much power compared to the observational measurements.  In the top right panel of Fig.~\ref{fig:shear_CMB_powerspec} we examine the cosmology dependence of the cross-spectrum.  Consistent with the previous statistics explored above, we find that lowering the value of $S_8$ (whether directly, or via an increase in the summed neutrino mass) improves the match to the measurements, with the LS8 cosmology yielding the best match.  Even the LS8 cosmology is somewhat elevated with respect to the measurements, however.  This is consistent with the findings of \citet{Robertson2021}, who used linear theory + \textsc{halofit} \citep{Smith2003,Takahashi2012} to predict the cross-spectrum and measured $S_8 = 0.64 \pm 0.08$.  A more recent configuration-space analysis by \citet{Chang2023}, using galaxy lensing and galaxy clustering data from DES Y3 cross-correlated with Planck + SPT CMB lensing, found a somewhat larger amplitude of $S_8 \approx 0.73 \pm 0.03$, though still low compared to the D3A or Planck CMB cosmologies.  For comparison, the D3A, Planck, and LS8 cosmologies have S$_8$ values of 0.815, 0.833, 0.766, respectively.

The bottom panels of Fig.~\ref{fig:shear_CMB_powerspec} examine the feedback dependence of the predicted shear--CMB lensing cross.  Minor effects are visible at $\ell \ga 500$ but these do not alter the conclusions drawn above.  

Going forward, there is considerable promise in measurements of this particular cross-spectrum, as high signal-to-noise tomographic analysis will be possible by combining Euclid and LSST shear measurements with Planck, Advanced ACT, and Simons Observatory CMB lensing measurements, allowing one to bridge the gap in scale and redshift between current cosmic shear-only and CMB lensing-only measurements.  

\subsection{Summary of comparison to observational data}

We provide here an overall summary of the comparison of the \flamingo\ suite of simulations to the various observed auto- and cross-spectra discussed above, along with the CMB lensing--tSZ effect cross-spectrum presented in \citet{Schaye2023}.  In particular, in Table \ref{tab:chi2_nsig} we present the $\chi^2$ values and number of standard deviations, $N_\sigma$, of each simulation with respect to the observed power spectra.  For the cosmic shear power spectrum ($\gamma_\textrm{E}$-$\gamma_\textrm{E}$) and cosmic shear--tSZ effect cross-spectrum ($\gamma_\textrm{E}$-$y$), we use the full covariance matrices to compute $\chi^2$, whereas for the other cases we use the diagonal elements of the covariance matrix only, as the covariance matrices were either unavailable or could not be straightforwardly combined when multiple data sets were used.  The numbers displayed in bold face correspond to the models that are within $3\sigma$ of the observational measurements.  

We see from Table \ref{tab:chi2_nsig} that the LS8 model performs the best for all the comparisons apart from the CMB lensing power spectrum, where it is clearly disfavoured, and the DES Y3 cosmic shear power spectrum (PlanckNu0p24Fix performs marginally better).  However, the cosmic shear--CMB lensing ($\gamma_\textrm{E}$-$\kappa_\textrm{CMB}$) and CMB lensing--tSZ effect\footnote{Just prior to submission of this paper, \citet{McCarthy2023} presented an updated measurement of the CMB lensing-tSZ effect cross-spectrum using Planck data.  For their highest signal-to-noise measurement, they find an amplitude of $A=0.82 \pm 0.21$ compared to a Planck cosmology with expectation of $A=1$.  \citet{Hill2014} found $A=1.10\pm0.22$ with respect to the same cosmology.  In the analysis of \citet{Schaye2023} and in Table \ref{tab:chi2_nsig}, we scaled the amplitude of \citet{Hill2014} down by 20\% based on \citet{Hurier2015}, which would correspond to $A=0.92\pm0.18$, which is consistent with the new findings of \citet{McCarthy2023}.  The slightly lower amplitude of the new measurements implies a slightly stronger tension with the Planck cosmology, though only at the $\approx1\sigma$ level (compared to $0.6\sigma$ in Table \ref{tab:chi2_nsig}).} ($\kappa_\textrm{CMB}$-$y$) cross-spectra are not particularly constraining at present and the Planck and D3A cosmologies are only in mild tension with the data for these tests.  The strongest evidence for a $S_8$ tension comes from the tSZ effect power spectrum ($y$-$y$), the cosmic shear--tSZ effect cross-spectrum ($\gamma_\textrm{E}$-$y$) and the cosmic shear power spectrum ($\gamma_\textrm{E}$-$\gamma_\textrm{E}$) in order of decreasing tension level.  This highlights the importance of the tSZ effect for constraining cosmology.  We also highlight that when the small-scale SPT measurements are included in the tSZ power spectrum comparison, all models are formally rejected.  As noted previously, some combination of a somewhat reduced value of $S_8$ and more extreme feedback may be able to accommodate the SPT measurements, but we leave that for future work.

While Table \ref{tab:chi2_nsig} provides a convenient way to compare the models and to demonstrate which tests are most constraining, there are two caveats to bear in mind, particularly with regard to the level of tension reported.  First, we have not marginalised over any possible sources of systematic error in the observations, such as photometric redshift uncertainties and intrinsic alignments in the case of cosmic shear, CIB leakage in the tSZ effect, and so on.  Indeed, this likely explains why the LS8 model does not provide a particularly good fit to the KiDS 1000 cosmic shear power spectrum, with $N_\sigma = 2.6$ in that case (although we note that approximately $1\sigma$ of this tension comes from a single cross, the 5-1 bin, which shows large point-to-point fluctuations). Marginalising over these uncertainties will tend to decrease the magnitude of the reported deviations.  Secondly, we quote the $\chi^2$ and $N_\sigma$ values with respect to a specific set of cosmological parameters.  For example, for the Planck simulation we use the Planck 2018 maximum-likelihood cosmological parameters and the statistics we report are therefore with respect to that specific choice of parameters.  The level of tension we report does not factor in the uncertainties in the cosmological parameters from the Planck primary CMB data set, it is only with respect to the maximum-likelihood cosmology.  Thus, our estimates will be an upper limit for the level of tension between the primary CMB and LSS observables in this case.  

\section{Discussion and Conclusions}
\label{sec:conclusions}

We have used the new \flamingo\ simulations to explore the spatial clustering and cross-correlation signals of three large-scale tracers: cosmic shear, CMB lensing, and the thermal Sunyaev-Zel'dovich (tSZ) effect.  Our analysis, carried out in harmonic space, included an exploration of the dependence of these signals on the choice of cosmological parameters including neutrino mass, the efficiency and nature of so-called `baryonic feedback' (specifically stellar and AGN feedback), and simulation box size, resolution, and method of analysis (1D Limber integration vs.~a 2D map-based analysis).  Note that the stellar and AGN feedback were calibrated using machine learning methods to reproduce the observed galaxy stellar mass function and the gas fractions of galaxy groups and clusters.  However, \flamingo\ also includes extreme variations with respect to the observed baryon fractions, obtained by varying the feedback efficiencies.  We compared the full \flamingo\ suite of hydro simulations to the latest measurements of the auto- and cross-power spectra involving the three observables and commented on the evidence for an `$S_8$ tension' and its robustness to uncertainties in baryonic modelling.

The main conclusions of this study may be summarised as follows:
\begin{itemize}
    \item The auto- and cross-power spectra involving cosmic shear, CMB lensing, and the tSZ effect include contributions from a very wide range of physical scales and redshifts (Fig.~\ref{fig:keff_zeff}).  While there is considerable overlap between the different 2-point functions in terms of physical scale and redshift (allowing for consistency tests), there is also a large degree of complementarity, with the CMB lensing power spectrum probing generally linear scales and higher redshifts while the cosmic shear and tSZ effect power spectra probe low redshifts and into the non-linear regime though they are sensitive to different halo masses.  Cross-spectra between the different observables are typically sensitive to physical scales, redshifts, and halo masses that are intermediate between the power spectra of the individual fields, thus adding additional information.
    \item The predicted cosmic shear power spectrum is robust to variations in box size, resolution, and method of analysis (Fig.~\ref{fig:shear_powerspec_1}).  Consistent with recent studies (e.g., \citealt{Heymans2021,Troster2022,Amon2023}), we find that a Planck CMB cosmology, and the fiducial cosmology which combines DES Y3 constraints with a variety of external data sets including the primary CMB (D3A), predicts power spectra that are elevated with respect to the KiDS 1000 measurements and to a lesser extent the DES Y3 measurements.  This is in the context of our fiducial calibrated hydro simulation.  We find that adopting a lower value of $S_8$, as suggested by recent cosmic shear measurements (LS8 run), yields an improved match to the measurements, particularly for the higher tomographic bins which contain much of the integrated signal (Fig.~\ref{fig:shear_powerspec_2}).  Increasing the feedback efficiency with respect to the fiducial calibrated model, which has the effect of lowering the gas fractions of groups and clusters, also marginally improves the quality of the fit (Fig.~\ref{fig:shear_powerspec_3}) in the context of the D3A cosmology, but this preference largely evaporates when adopting a lower $S_8$.  Even though it contains extreme variations in feedback, \flamingo\ predicts that current cosmic shear measurements are only marginally sensitive to the impact of baryon physics. 
    \item The predicted tSZ effect power spectrum is generally robust to variations in resolution, box size, and method of analysis (Fig.~\ref{fig:tSZ_powerspec}).  As much of the signal is produced by local, massive clusters, very large simulation boxes ($\ga 1$ Gpc) are required to accurately predict the power.  Another consequence of this dependence on massive, nearby clusters is that cosmic variance uncertainties are important on large scales (particularly $\ell \la 300$).  The tSZ effect power spectrum is extremely sensitive to the amplitude of fluctuations ($\sigma_8$) and, on the angular scales probed by Planck, is generally insensitive to variations in the efficiency of feedback (owing to the fact that it is dominated by `baryonically-closed' massive clusters).  The situation changes on smaller angular scales probed by SPT and ACT, which are more sensitive to group-mass haloes and therefore more susceptible to feedback effects.  Current measurements of the tSZ effect on large scales \citep{Bolliet2018} are in strong tension with the predictions of the hydrodynamical simulations in either a D3A or Planck CMB cosmology (see Table \ref{tab:chi2_nsig}).  Adopting a lower value of $S_8$, however, yields a significantly improved match to the measurements, consistent with the cosmic shear power spectrum analysis.
    \item We find that the CMB lensing power spectrum is insensitive to simulation box size, resolution, method of analysis, and feedback implementation (see Fig.~\ref{fig:CMB_powerspec_feedback}).  This is a direct consequence of it probing mainly linear scales and high redshifts.  In contrast to the cosmic shear and tSZ effect power spectra, the predicted CMB lensing power spectrum for both the D3A and Planck CMB cosmologies is in excellent agreement with recent measurements from Planck, SPT, and ACT, whereas the cosmologies with an increased neutrino mass and (particularly) LS8 are disfavoured (Fig.~\ref{fig:CMB_powerspec}).
    \item The predicted cosmic shear--tSZ effect cross-spectrum is generally robust to variations in resolution, box size, and method of analysis (Fig.~\ref{fig:tSZ_shear_powerspec}).  Similar to the tSZ effect power spectrum, much of the signal is produced by local, massive clusters, thus requiring 
    large simulation boxes to accurately predict this statistic.  Consistent with \citet{Troster2022}, we find that the predicted cross-spectrum in both the D3A or Planck CMB cosmologies is elevated with respect to the observed cross-spectrum between KiDS 1000 and Planck and ACT tSZ on scales of $\ell \la 1500$.  Decreasing the gas fractions of groups and clusters by increasing the efficiency of feedback leads to a marginal improvement in the fit but cannot accommodate the observed offset on large angular scales ($\ell \la 1000$).  Adopting a lower $S_8$ improves the match to the data.  
    \item The predicted cosmic shear--CMB lensing cross-spectrum is very robust to variations in resolution, box size, and method of analysis (Fig.~\ref{fig:shear_CMB_powerspec}).  The impact of baryons is also minimal compared to current measurement uncertainties.  Consistent with \citet{Robertson2021}, we find that the predicted cross-spectrum in either the D3A or Planck CMB cosmologies is generally elevated with respect to the observed cross-spectrum between KiDS 1000 and Planck and ACT CMB lensing.  Adopting a lower $S_8$ improves the match to the data.
    \item We summarise the level of tension of the simulations with respect to the observations for all of the auto- and cross-power spectra we have considered in Table \ref{tab:chi2_nsig}.
\end{itemize}

Our findings are largely consistent with those of the individual studies which presented the measurements we compare to, at least in terms of qualitative conclusions.  This was not guaranteed to be the case, as the individual studies all use very different theoretical frameworks, which is a crucial part of cosmological parameter inference.  For example, some of the studies employ linear theory (e.g., \citealt{Madhavacheril2023,Qu2023}) while others use the halo model (e.g., \citealt{Bolliet2018,Troster2022}) or gravity-only emulators (e.g., \citealt{Robertson2021,Doux2022}) to compute the non-linear evolution.  With regards to the potential impact of baryons, in some studies it was not taken into account (e.g., \citealt{Robertson2021,Madhavacheril2023,Qu2023}) while others employed the halo model and marginalised over the uncertainties (e.g., \citealt{Bolliet2018,Troster2022}) while still others discarded small scale measurements (e.g., \citealt{Doux2022}) in an attempt to avoid biasing due to baryons.  Here we have taken a single suite of hydro simulations and projected them into different observables, thus allowing a direct comparison between the different tests.  From this comparison, three major conclusions are: i) power spectra that probe late times and/or non-linear scales are in tension with the predictions of the standard LCDM model with parameters set by the primary CMB, BAO, and CMB lensing; ii) while increasing the neutrino mass can reduce the tension between the CMB+BAO and LSS, measurements of the CMB lensing power spectrum disfavour this solution; and iii) the effects of baryon physics and, importantly, its uncertainties, are generally insufficient to reconcile these tensions.  

Our results do not preclude baryons from playing some minor role in the current discussion which, perhaps when combined with other factors (e.g., systematic errors in observational measurements), could help to resolve the tension.  Indeed recent cosmic shear analyses claim to have marginally detected deviations from the predictions of gravity-only simulations and at a level that is consistent with that predicted by calibrated simulations such as BAHAMAS and \flamingo\ (e.g., \citealt{Chen2023,Arico2023}).  And while including baryons in the modelling tends to boost the inferred value of $S_8$, in general the value is shifted by less than $1\sigma$.  Thus, both calibrated simulations and current cosmic shear measurements point to a fairly benign role for baryons at present.  However, this situation will change radically for forthcoming Stage IV surveys, including LSST, Euclid, and DESI, since the statistical precision of the measurements will be significantly higher and the measurements will extend to smaller scales, thus requiring a much more careful accounting of the impact of baryon physics.  

Returning to the recent studies of \citet{Amon2022} and \citet{Preston2023}, these authors showed that, given sufficient flexibility in the incorporated baryonic modelling, it is possible to obtain a satisfactory fit to both the primary CMB and cosmic shear measurements.  The required suppression of the matter power spectrum is, however, typically much stronger than predicted by simulations such as \bahamas\ and \flamingo, for which the baryon physics has been calibrated to reproduce the baryon fractions of galaxy groups and clusters.  Recent work has shown that the suppression is strongly tied to the baryon fractions (e.g., \citealt{Semboloni2011,Semboloni2013,Schneider2019,VanDaalen2020,Salcido2023}), which was a motivating factor in the calibration strategies of these simulations.  Thus, appealing to a much stronger suppression of the matter power spectrum in order to reconcile the primary CMB and LSS would appear to require baryon fractions that violate observational constraints on the baryon fractions of groups and clusters.  \citet{Grandis2023} have recently come to similar conclusions using weak lensing-calibrated cluster gas and stellar fractions together with the baryonification methods of \citet{Schneider2019} and \citet{Arico2021emulator} to predict the suppression of the matter power spectrum.  However, we note that the level of feedback required to reconcile the primary CMB and the cosmic shear measurements depends on the cosmic shear data set employed. \citet{Preston2023} find that relatively weaker feedback, which is more comparable to that in \bahamas\ and \flamingo, is required to reconcile the DES Y3 measurements with the CMB, whereas stronger feedback is required when using the KiDS 1000 measurements.  This is likely driven by the fact that the tension in $S_8$ between the DES Y3 measurements and the CMB is relatively mild to begin with, before baryonic effects are considered.

In terms of using baryon fractions to constrain baryonic effects on the matter power spectrum, a caveat that is worth further consideration is that carefully accounting for the X-ray selection function of galaxy groups is non-trivial (e.g., \citealt{Pearson2017}).  Also, the present cosmic shear data is sensitive to LSS over a wider range of redshifts than for which we presently have useful observational constraints on group/cluster baryon fractions.  Thus, if the simulations significantly underestimate the efficiency of feedback at higher redshifts ($z \ga 0.3$), then they could underestimate the impact of baryons on the cosmic shear power spectrum.  Observations of groups/clusters at higher redshift and with a well-defined selection function would be highly valuable.  The recent kinetic SZ (kSZ) effect stacking measurements of SDSS eBOSS galaxies at $z\approx 0.5$ in \citet{Schaan2021} appear to be well suited as an independent test of feedback models (e.g., \citealt{Schneider2022}).  We emphasise, however, that the most strongly discrepant power spectrum we examined is the Planck tSZ effect power spectrum, which is primarily sensitive to low-redshift, massive clusters which are baryonically-closed, dominated by hot gas, and for which X-ray and tSZ surveys are typically highly complete.  Increasing feedback at higher redshifts will not significantly impact this metric.

If neither feedback nor unaccounted for (or mischaracterised) systematic errors are behind the tension (though it will require additional work to conclusively demonstrate the latter), then the exciting implication would appear to be that new physics, perhaps in the dark sector, is required.  Specifically, new physics that preferentially impacts non-linear scales and/or late times in order to retain consistency with the CMB lensing power spectrum, as proposed by \citet{Amon2022} and \citet{Preston2023} as an alternative interpretation (to baryon physics) of the required suppression.  Many suggestions in this vein have recently been put forward, including a contribution from ultra-light axions to the dark matter (e.g., \citealt{Rogers2023}), mild baryon--dark matter scattering \citep{He2023}, and invoking interactions (in the form of a frictional drag) between dark matter and dark energy \citep{Poulin2023}.  Testing these extensions will require forthcoming Stage IV surveys, which will measure the power and cross-power spectra with much higher statistical precision and will allow for much finer binning in redshift, physical scale, and halo mass.  With the increased statistical precision comes the requirement for a very careful consideration of baryonic effects, which may be degenerate with the additional degrees of freedom in the dark sector extensions.

Finally, to make stronger statements about tensions and possible resolutions thereof in the context of cosmological hydrodynamical simulations and which simultaneously accounts for relevant observational systematic uncertainties, we will require a way to quickly span a wide range of baryon feedback scenarios and background cosmologies in order to incorporate the predictions of such simulations into cosmological pipelines.  The latter (cosmological variations) has already been achieved but only in the context of gravity-only simulations, through emulators constructed from grids of simulations that span some range of cosmological parameters (e.g., \citealt{Heitmann2014,Lawrence2017,Euclid2019,McClintock2019,Euclid2021}).  The former has just recently been achieved for the first time in volumes of sufficient size for LSS applications, in \citet{Salcido2023}.  Those authors produced a suite of 400 cosmological hydro simulations in 100 Mpc$/h$ volumes (the Antilles simulations) and have developed an emulator for the relative impact of baryons on the matter power spectrum that takes the mass-dependent baryon fractions of groups and clusters as its input and predicts the suppression of the matter power spectrum.  In the context of current cosmic shear measurements, this is the most accurate emulator currently available and can be easily applied in existing pipelines.  

However, for the next generation of measurements, which will have significantly improved statistical precision, the assumed separability of baryonic effects and cosmological effects in the approach of \citet{Salcido2023} and other approximate methods for incorporating baryons (e.g., HMcode; \citealt{Mead2020}) will need to be revisited.  Ideally, a single emulator based on a grid of cosmological hydrodynamical simulations that \textit{simultaneously} varies the relevant astrophysical and cosmological parameters should be the basis of cosmological pipelines.  Furthermore, to take advantage of the wide variety of complementary LSS observables, including weak lensing, galaxy clustering, redshift-space distortions, the tSZ and kSZ effects, cluster counts, etc. box sizes of $\sim 1$ Gpc are required.  With \flamingo\ we have taken an important step forward to show that it is possible to carry out such simulations in a careful way, using machine learning-based emulators as part of the calibration.  The next step is to extend this approach to a simultaneous exploration of cosmology and astrophysics, which is the subject of ongoing work.

\section*{Acknowledgements}
The authors thank the referee for a prompt and constructive report.
IGM thanks Tilman Tr{\"o}ster and Cyrille Doux for assistance with KiDS and DES cosmic shear measurements and Alexandra Amon for helpful comments on a draft version of the paper.
This project has received funding from the European Research Council (ERC) under the European Union’s Horizon 2020 research and innovation programme (grant agreement No 769130). CSF acknowledges support from the European Research Council (ERC) through Advanced Investigator grant DMIDAS (GA 786910).  This work used the DiRAC@Durham facility managed by the Institute for Computational Cosmology on behalf of the STFC DiRAC HPC Facility (\url{www.dirac.ac.uk}). The equipment was funded by BEIS capital funding via STFC capital grants ST/K00042X/1, ST/P002293/1, ST/R002371/1 and ST/S002502/1, Durham University and STFC operations grant ST/R000832/1. DiRAC is part of the National e-Infrastructure.  

%%%%%%%%%%%%%%%%%%%%%%%%%%%%%%%%%%%%%%%%%%%%%%%%%%
\section*{Data Availability}

The data supporting the plots within this article are available on reasonable request to the corresponding author. The \flamingo\ simulation data will eventually be made publicly available, though we note that the data volume (several petabytes) may prohibit us from simply placing the raw data on a server. In the meantime, people interested in using the simulations are encouraged to contact the corresponding author.

%%%%%%%%%%%%%%%%%%%% REFERENCES %%%%%%%%%%%%%%%%%%

% The best way to enter references is to use BibTeX:

\bibliographystyle{mnras}
\bibliography{references} % if your bibtex file is called example.bib

%%%%%%%%%%%%%%%%%%%%%%%%%%%%%%%%%%%%%%%%%%%%%%%%%%

%%%%%%%%%%%%%%%%% APPENDICES %%%%%%%%%%%%%%%%%%%%%

\appendix

\section{Additional comparisons}
\label{sec:other}

In this Appendix we present additional analyses referred to in the main text.

\begin{figure*}
    \includegraphics[clip, trim=3.0cm 2.5cm 3.0cm 4.0cm, width=\textwidth]{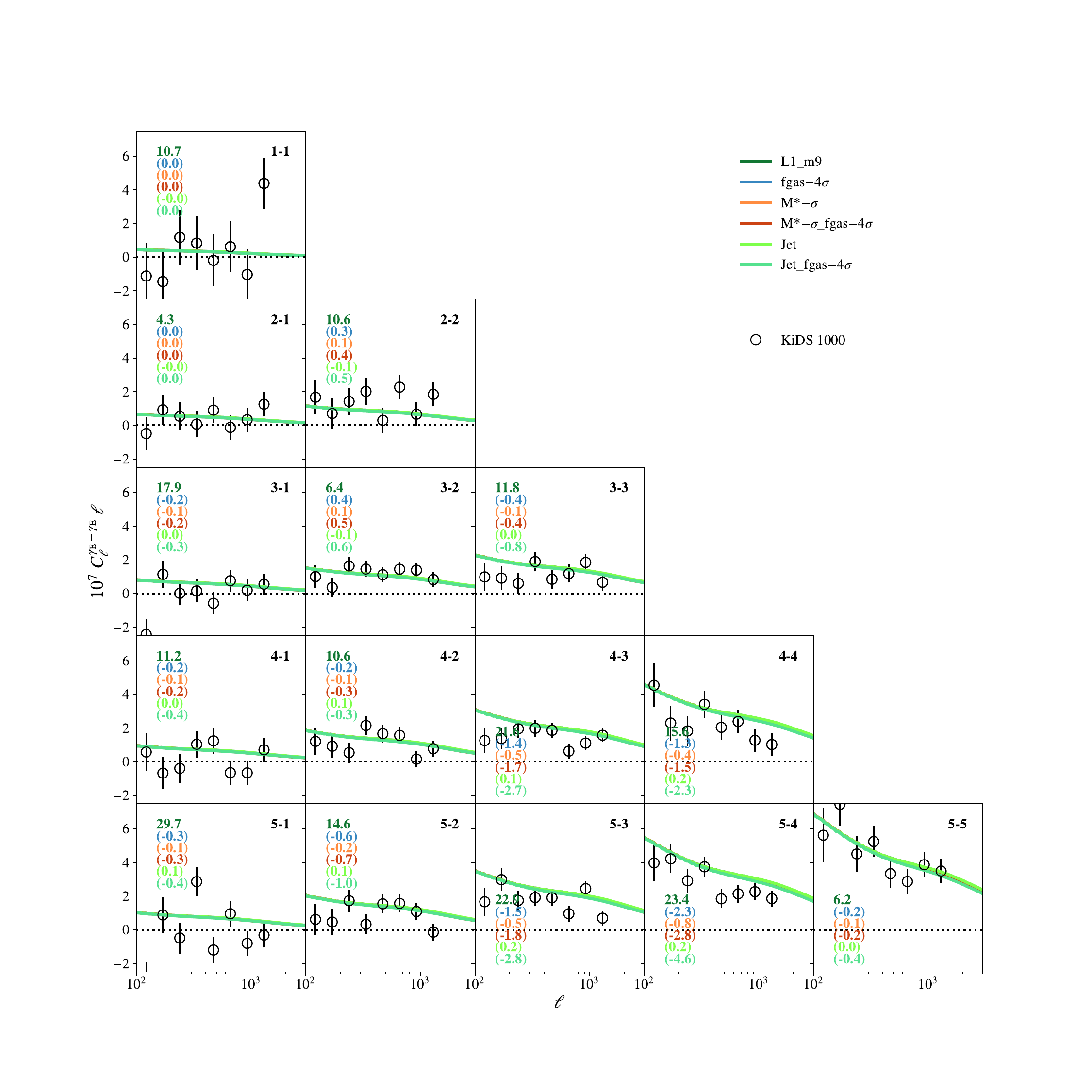}
    \caption{As Fig.~\ref{fig:shear_powerspec_1}, but showing the dependence of the predicted KiDS 1000 cosmic shear power spectrum on baryon physics, namely variations in the stellar mass function (both the fiducial and reduced cluster gas fractions) and the fiducial and strong jet models of AGN feedback.  The solid coloured curves correspond to the predicted spectra for the \flamingo\ simulations as baryon models are varied.  The effect of variations in the stellar mass function is generally negligible compared to that of variations in the gas fractions.}
    \label{fig:shear_powerspec_4}
\end{figure*}

In Fig.~\ref{fig:shear_powerspec_4} we examine the dependence of the predicted KiDS 1000 cosmic shear power spectrum on baryon physics, specifically variations in the SMF (both at the fiducial and reduced gas fractions) and the fiducial and strong jet models of AGN feedback.  We conclude that variations in the SMF are generally negligible compared to that of variations in the gas fractions.

\begin{figure*}
    \includegraphics[clip, trim=2.0cm 2.0cm 2.0cm 2.0cm, width=\textwidth]{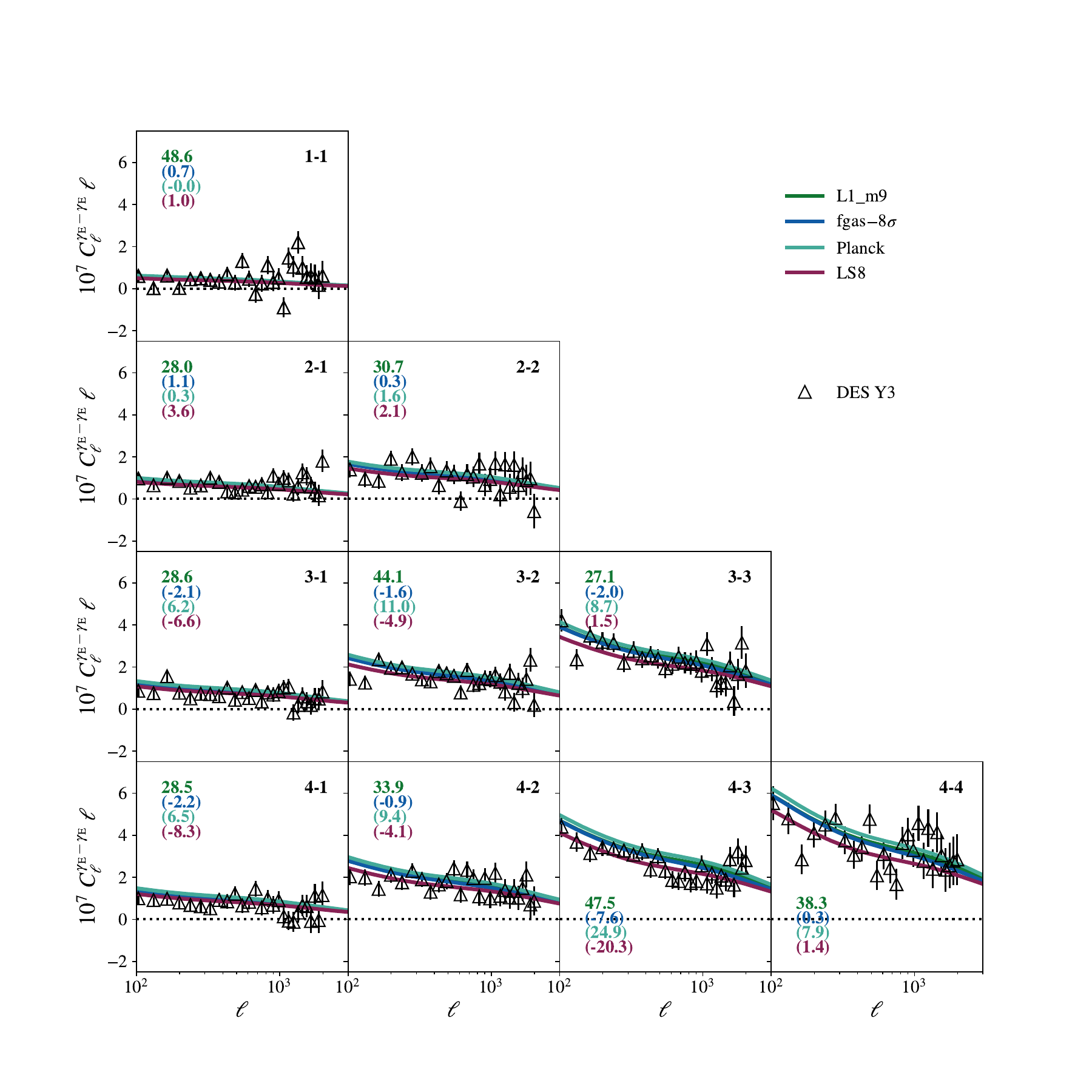}
    \caption{As Fig.~\ref{fig:shear_powerspec_1}, but showing selected predictions for the DES Y3 cosmic shear power spectrum.    The open triangles correspond to the DES Y3 measurements of \citet{Doux2022} and the error bars correspond to the diagonal components of the covariance matrix.  Note that the $\chi^2$'s are computed here adopting the diagonal errors only.  A lensing LS8 cosmology yields somewhat better fit to the data relative to the fiducial D3A cosmology, whereas a Planck CMB cosmology yields a worse fit.  Increasing the efficiency of feedback also slightly improves the fit, but less than the improvement due to lowering $S_8$.}
    \label{fig:shear_powerspec_DES}
\end{figure*}

In Fig.~\ref{fig:shear_powerspec_DES} we compare selected \flamingo\ runs to the DES Y3 cosmic shear power spectra data from \citet{Doux2022}.  The DES Y3 background galaxy population is split amongst 4 tomographic bins shown in figure 2 of \citet{Doux2022}.  We use these source redshift distributions in eqn.~\ref{equ:W-gamma} to compute their respective window functions.  As we are particularly interested in the role of baryons, we do not include any scale cuts to the DES Y3 power spectra.  Examining Fig.~\ref{fig:shear_powerspec_DES}, we see that the lensing LS8 cosmology yields a somewhat better fit to the data relative to the fiducial D3A cosmology (particularly amongst the majority of the higher tomographic bins), whereas a Planck CMB cosmology yields a worse fit for virtually all bins.  Increasing the efficiency of feedback also slightly improves the fit, but less than the improvement due to lowering $S_8$.  Consistent with findings in the literature (e.g., \citealt{Abbott2022,Abbott2023}), we conclude that while there is some evidence for tension of the DES Y3 measurements with the Planck CMB cosmology, it is of slightly lower significance than for the KiDS 1000 survey.

\begin{figure}
    \includegraphics[width=\columnwidth,left]{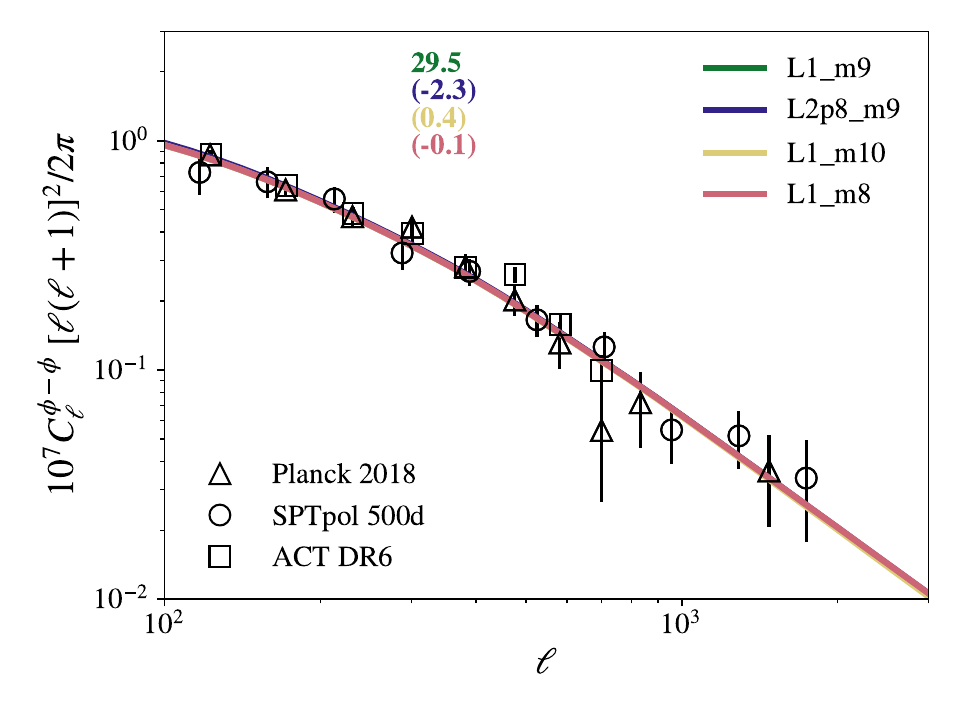}
    \includegraphics[width=\columnwidth]{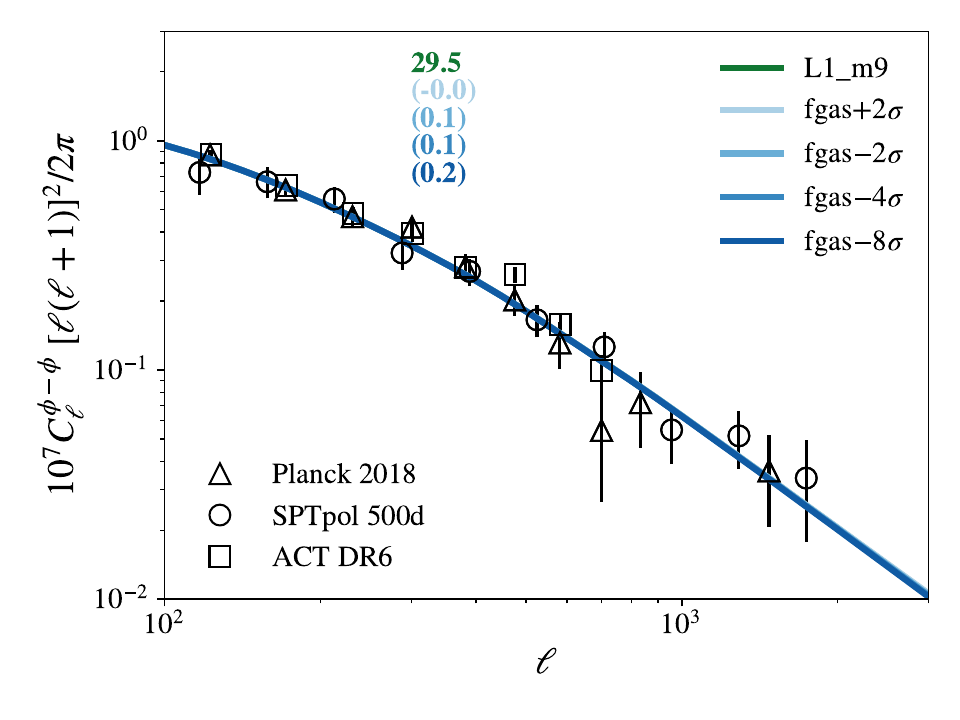}
    \includegraphics[width=\columnwidth]{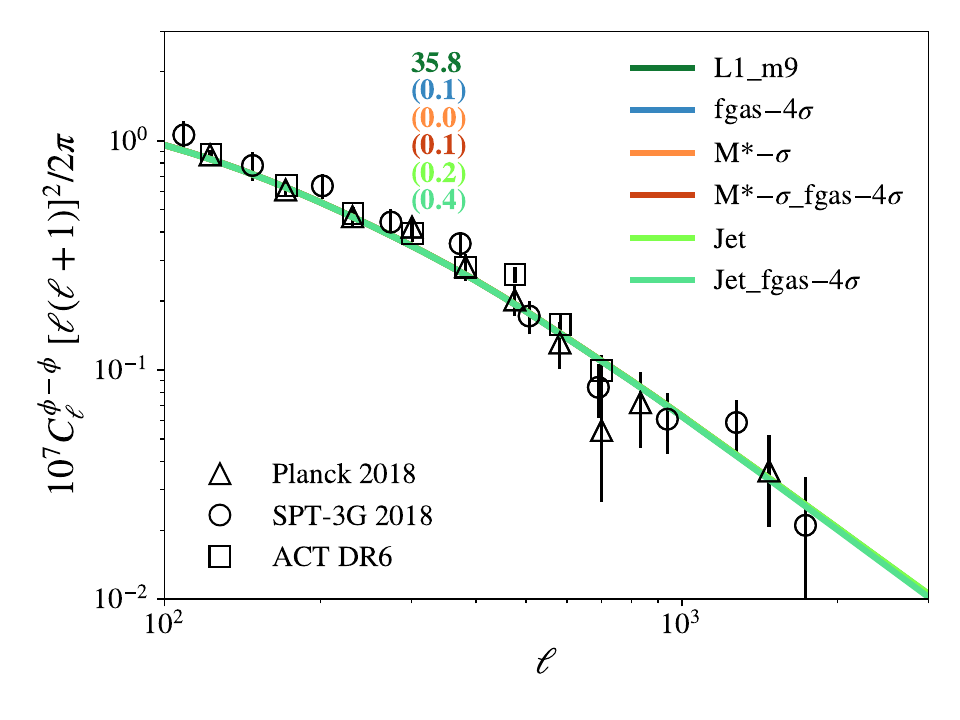}
    \caption{As Fig.~\ref{fig:CMB_powerspec}, but showing the dependence of the predicted CMB lensing angular power spectrum on simulation box size and resolution (top panel) and baryon physics (middle and bottom panels).  Over the range of scales examined here, the predicted CMB lensing power spectrum is converged (i.e., unaltered) with respect to these variations.}
    \label{fig:CMB_powerspec_feedback}
\end{figure}

In Fig.~\ref{fig:CMB_powerspec_feedback} we examine the dependence of the predicted CMB lensing angular power spectrum on simulation box size and resolution (top panel) and baryon physics (bottom panels).  Over the range of scales examined here, the predicted CMB lensing power spectrum is converged (i.e., unaltered) with respect to these variations.

\begin{figure}
    \includegraphics[width=\columnwidth]{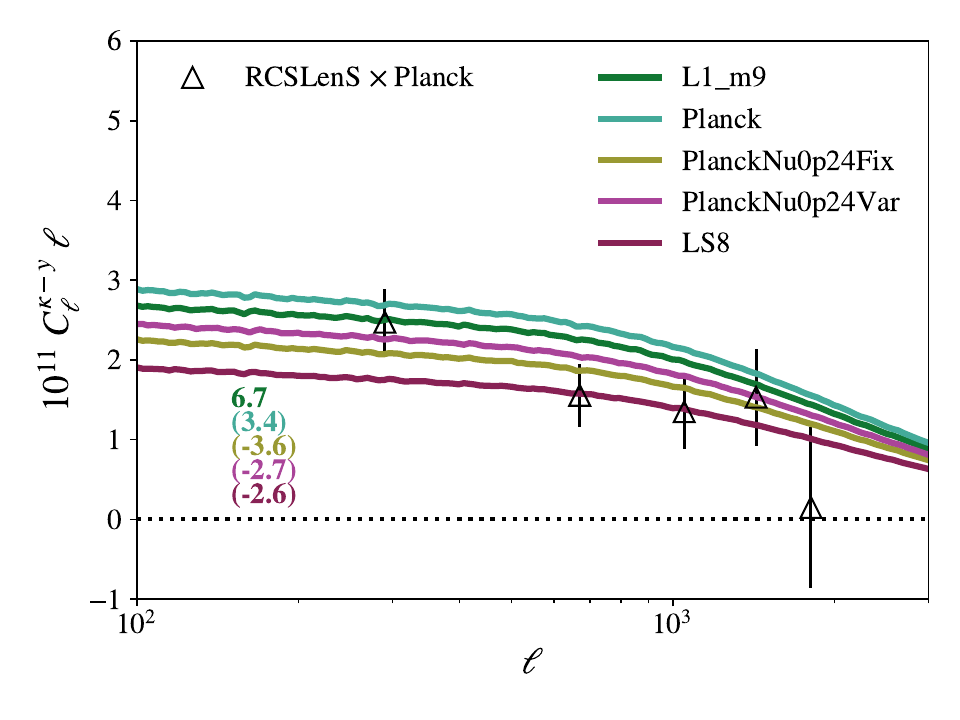}\\
    \includegraphics[width=\columnwidth]{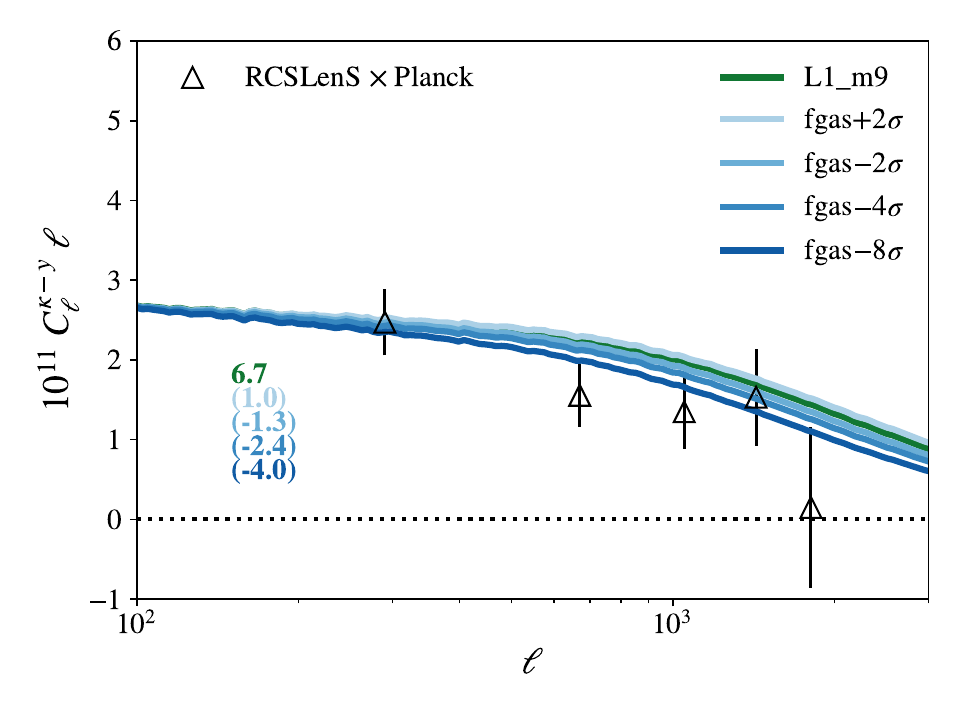}
    \caption{The cosmic shear--tSZ effect angular cross-power spectrum.  The open triangles correspond to the RCSLenS $\times$ Planck tSZ measurements of \citet{Hojjati2017}. \textit{Top:} Dependence on cosmology.  The solid coloured curves correspond to the predicted spectra for the \flamingo\ simulations for different background cosmologies. \textit{Bottom:} Dependence on baryon physics, namely variations in the gas fractions of groups and clusters, which are mediated primarily through variations in the AGN feedback strength.  The solid coloured curves correspond to the predicted spectra for the \flamingo\ simulations as the gas fractions are varied from $+2\sigma$ to $-8\sigma$ with respect to the observed gas fraction--halo mass relation.}
    \label{fig:tSZ_shear_RCS_powerspec}
\end{figure}

In Fig.~\ref{fig:tSZ_shear_RCS_powerspec} we compare the \flamingo\ cosmology and gas fraction variation runs with the cosmic shear--tSZ effect angular cross-power spectrum measurements of \citet{Hojjati2017}, using data from RCSLenS (shear) and Planck (tSZ).  As RCSLenS is a relatively shallow survey, \citet{Hojjati2017} combined the data into a single large tomographic bin.  We use the source redshift distribution from that study to compute the predicted shear--tSZ effect cross.  Consistent with the KiDS 1000 $\times$ Planck cross examined in the main text, we see that four of the five multipole bins have amplitudes lower than that predicted by the fiducial \flamingo\ model in the fiducial D3A cosmology.  However, as these four bins sample relatively small angular scales of $\ell \sim 1000$, where the impact of feedback is non-negligible, the significance of the tension with the D3A cosmology is clearly less pronounced than for the KiDS 1000 comparison in the main text.

%%%%%%%%%%%%%%%%%%%%%%%%%%%%%%%%%%%%%%%%%%%%%%%%%%

% Don't change these lines
\bsp	% typesetting comment
\label{lastpage}
\end{document}